\documentclass[%
 aip,
 amsmath,amssymb,
 reprint,%
]{revtex4-1}

\usepackage{siunitx}
\usepackage{graphicx}
\usepackage{dcolumn}
\usepackage{bm}

\usepackage[utf8]{inputenc}
\usepackage[T1]{fontenc}
\usepackage{mathptmx}
\usepackage{etoolbox}
\usepackage{booktabs}

\usepackage[normalem]{ulem}

\makeatletter
\def\@email#1#2{%
 \endgroup
 \patchcmd{\titleblock@produce}
  {\frontmatter@RRAPformat}
  {\frontmatter@RRAPformat{\produce@RRAP{*#1\href{mailto:#2}{#2}}}\frontmatter@RRAPformat}
  {}{}
}%
\makeatother
\begin{document}


\title{Single-LED-pumped, room-temperature, solid-state maser\\}


\author{Michael Newns}
\affiliation{Dept.~of Materials, Imperial College London, South Kensington, SW7 2AZ, UK}%
\author{Shirley Xu}%
\affiliation{Dept.~of Materials, Imperial College London, South Kensington, SW7 2AZ, UK}%
\author{Mingyang Liu}%
\affiliation{Dept.~of Physics, Imperial College London, South Kensington, SW7 2AZ, UK}%
\author{Yifan Yu}%
\affiliation{Dept.~of Materials, Imperial College London, South Kensington, SW7 2AZ, UK}%
\author{Zike Cheng}%
\affiliation{Dept.~of Materials, Imperial College London, South Kensington, SW7 2AZ, UK}%
\author{Vanessa Ussalim}%
\affiliation{Dept.~of Materials, Imperial College London, South Kensington, SW7 2AZ, UK}%
\author{Zitong Qin}%
\affiliation{Dept.~of Materials, Imperial College London, South Kensington, SW7 2AZ, UK}%
\author{Ziqiu Huang}%
\affiliation{Dept.~of Materials, Imperial College London, South Kensington, SW7 2AZ, UK}%
\author{Max Attwood}
\affiliation{Dept.~of Materials, Imperial College London, South Kensington, SW7 2AZ, UK}%
\author{Mark Oxborrow}
\affiliation{Dept.~of Materials, Imperial College London, South Kensington, SW7 2AZ, UK}
\email{m.oxborrow@imperial.ac.uk}


\begin{abstract}
Optically-pumped, solid-state (OPSS) masers show promise as quantum sensors, oscillators, and amplifiers --even when operating at room temperature. 
Here we show that the brightness (\textit{i.e.}, the luminance emitted at the surface) of a light-emitting diode (LED) is sufficient to pump such a maser above threshold. 
Specifically, we demonstrate pulsed maser oscillation in a microwave cavity
at room temperature containing a crystal of pentacene-doped
\textit{para}-terphenyl (ptc:ptp) pumped by a single, commercial, chip-scale LED emitting green-yellow light. 
The LED is driven by discharging a physically small, low-voltage capacitor, resulting in a bench-top maser that
is more compact, and a good deal cheaper to make, than existing embodiments. 
The ptc:ptp crystal is pumped invasively through a fused-quartz waveguide whose outputting facets
are shaped like the tip of a spear. An alternative butt-coupled geometry was also tried but failed to mase.
Towards explaining our results, the established mean-field formula for a maser's cooperativity is extended to account for
a spatially varying optical pump intensity (thus excitation rate) across a spatially varying a.c.~magnetic field $\textbf{B}(\vec{\textbf{r}})$ (of the cavity's TE$_{01\delta}$ maser mode). This extension leads to a dimensionless figure of merit, we call the ``strike fraction'' (appearing as a factor in the cooperativity), that quantifies how well the applied optical pump targets
those regions of the maser crystal where $\textbf{B}(\vec{\textbf{r}})$ is strongest.
We go on to show how, provided the spatial profile of $\textbf{B}(\vec{\textbf{r}})$ is known (through electromagnetic FEM-based simulation or otherwise), the strike fraction can be straightforwardedly  estimated, via Monte Carlo integration, from the data set of an optical ray-tracing simulation of the pump light. 
These developments allow a Tavis-Cummings-like model (master equation), encompassing terms up to 2$^{\textrm{nd}}$-order in quantum cumulants,
to simulate the process of masing to a greater level of realism than has hitherto been implemented. 
Provided the power of the optical pump is not a limiting resource, we find that the spear-head geometry offers at least a factor-of-two improvement (in terms of the maser's cooperativity) over butt-coupling. Furthermore, our ray-tracing simulations indicate that, even with invasing optical pumping, only a small fraction of the maser crystal's volume is adequately illuminated, suggesting that the cooperativity of optically-pumped masers based on concentrated (and thus optically dense) gain materials 
can still be substantially improved through better optical design. 
\end{abstract}

\maketitle

As quantum devices exploiting stimulated emission, room-temperature masers enable applications that cannot tolerate the severe operating constraints imposed by refrigeration. It is known that masers can amplify weak microwave signals with low noise figures\cite{clauss2008ruby}, detect extremely weak a.c.~magnetic fields\cite{wu_quantum_sensing2022}, and generate tones at GHz frequencies exhibiting low phase noise\cite{Goldenberg1960AtomicMaser,varshavsky2025solid}. It is also understood that, for these capabilities to persist at room-temperature, the maser must operate at high cooperativity such that the rate of stimulated emission greatly exceeds the rate of incoherent transitions associated with thermal noise~\cite{Oxborrow2012MaserAssembly,Sherman2022Diamond-basedAmplifier}. From a practical standpoint, room-temperature masers are still more cumbersome than semiconductor-based alternatives, due to the size and weight of the electromagnets and/or the light sources required for their operation. This remaining bulk obviates many of the advantages that follow from overcoming the need for refrigeration alone.

\begin{figure}[h]
    \centering
    \includegraphics[width=1\linewidth]{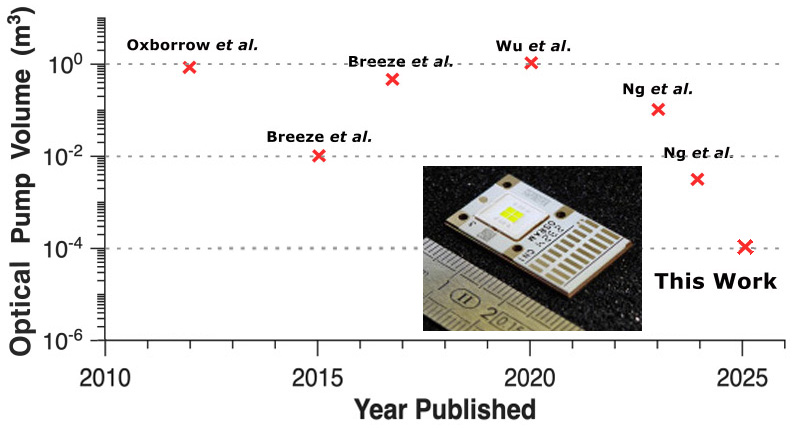}
    \caption{Approximate volume of optical pump source for key publications on ptc:ptp masers vs year published.  The inset photo displays the model of LED used in this work --beside a ruler for scale. References in ascending year are: Oxborrow \textit{et al.}~\cite{Oxborrow2012Room-temperatureMaser}, Breeze \textit{et al.}~\cite{Breeze2015EnhancedMasers}, Breeze \textit{et al.}~\cite{Breeze2017Room-temperatureStates}, Wu \textit{et al.}~\cite{Wu2020InvasiveCoupling}, Ng \textit{et al.}~\cite{Ng2023MoveElectrodynamics}, Ng \textit{et al.}~\cite{Ng2024Maser-in-a-shoebox:Field}.}
    \label{fig:pumpsize_vs_year}
\end{figure}

In the above context, the original series of room-temperature solid-state masers\cite{Oxborrow2012Room-temperatureMaser,Breeze2015EnhancedMasers,Wu2020InvasiveCoupling,Wu2020Room-TemperatureConcentrator},
based on optically-pumped pentacene-doped \textit{para}-terphenyl (ptc:ptp), remain attractive. Unlike masers based on other emitters such as NV centers in diamond, no strong applied DC magnetic field is needed. Furthermore, ptc:ptp masers can sustain high cooperativity in bursts lasting several milli-seconds~\cite{Wu2020Room-TemperatureConcentrator}, which is long enough to be useful in technologically important pulsed applications such as radar and spin-echo EPR. Such a zero-field maser\cite{bogle1959zero} is composed of just two essential parts: the maser resonator or cavity itself (including the maser crystal and the microwave mode it interacts with) and its associated optical pump source.
\begin{figure}[ht!]
    \centering
    \includegraphics[width=1\linewidth]
    {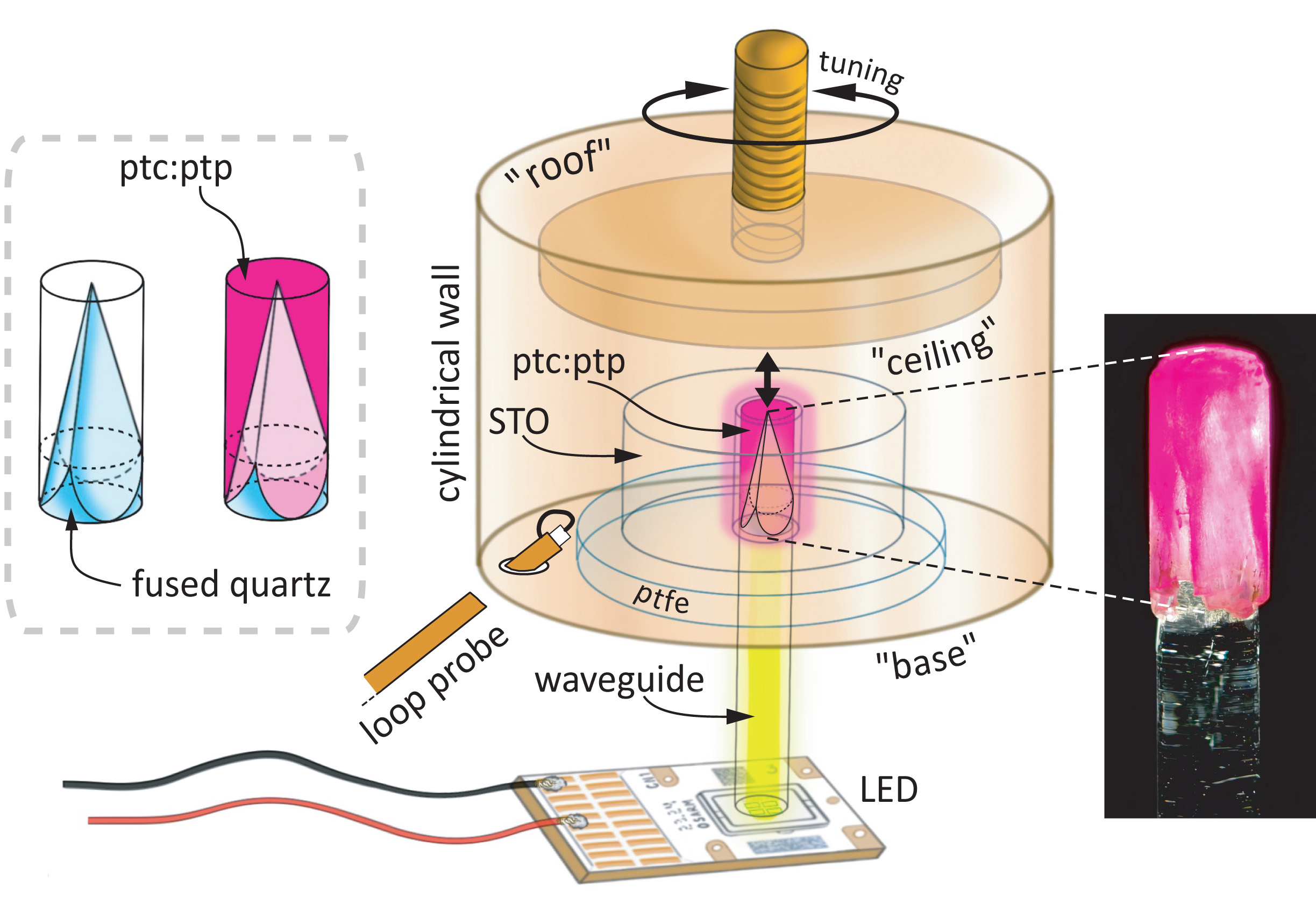}
    \caption{Geometry of the invasively pumped LED maser, including the LED chip, microwave cavity, waveguide and ptc:ptp crystal. The gray-dashed inset box on the the
    left details the geometry of the waveguide's 3-faceted tip. Displayed on the right is a photo of the invasively pumped ptc:ptp sample used in this work.}
    \label{fig:anatomy}
\end{figure}
Hitherto, the size, weight and power (`SWAP') requirements of a ptc:ptp maser have been dominated by the latter; typically the pump source has been at least an order of magnitude larger in volume than the maser cavity. Sources used (please see Fig.~\ref{fig:pumpsize_vs_year}) have included long-pulse liquid dye lasers\cite{Oxborrow2012Room-temperatureMaser}, various solid-state lasers (including OPOs)\cite{Breeze2015EnhancedMasers,Ng2024Maser-in-a-shoebox:Field}, and luminescence concentrators (both flash-lamp \cite{Wu2020Room-TemperatureConcentrator} and LED-pumped \cite{Sathian2025LED-pumpedMaser}), with volumes ranging from ``coffin'' to ``shoebox''. To be useful, ptc:ptp masers must offer improved performance over existing room-temperature technologies whilst offering SWAP metrics significantly lower than cryogenic alternatives. 

In this letter, we demonstrate pulsed maser action in ptc:ptp using invasive optical pumping from a single, commercially available, chip-scale green-yellow InGaN LED. This LED is driven by current from a capacitor, functioning as an energy store, whose volume dominates the size of the hardware required.[The pulsed current source's other components, like its MOSFET switch and control electronics are far smaller.] Because modern InGaN LEDs are several times more efficient at converting electrical power into pump photons than the above-mentioned laser- or LC-based pump sources, the required energy-storage capacitor can be physically smaller than the maser cavity itself. For our demonstrator, the total volume comes in at around 100 cm$^2$. As Fig.~\ref{fig:pumpsize_vs_year} displays, this represents a significant miniaturisation of the maser's overall size and weight. 
A notable omission from this figure is the work of ref.~\onlinecite{Sathian2025LED-pumpedMaser} which reports a ptc:ptp maser pumped by an LED-driven Ce:YAG LC, but does not provide sufficient information to estimate the volume of its LEDs or drive circuitry. We do know that this maser used 2120 LEDs, each with an output area of 1~mm~$\times$~1~mm, thus a total output area of 2120~mm$^2$. In comparison, this work used an LED output area of just 8.32~mm$^2$.

\section{Experimental Work}

\begin{figure}[h!]
    \centering
    \includegraphics[width=0.7\linewidth]
    {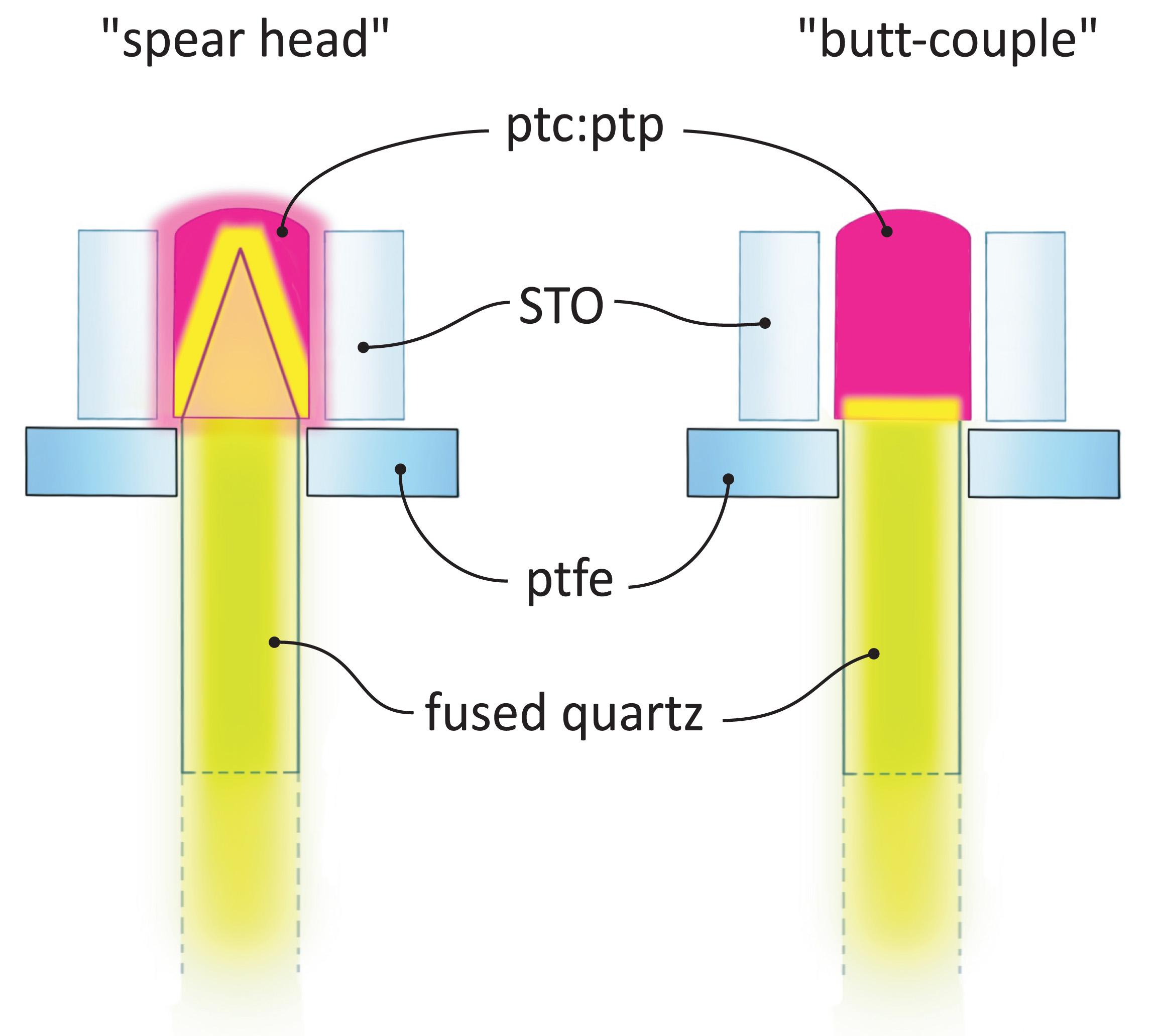}
    \caption{Meridional cross-sections through the optically-pumped maser's core for the two different pumping geometries considered. In both cases, the STO ring rests on a ptfe washer that itself rests on the cavity's copper base. Yellow shading indicates the extent of the pump light's penetration.}
    \label{fig:cross_section}
\end{figure}

In this work, a 0.1\%~ptc:ptp crystal grown atop a wedged fused-quartz waveguide is used as the maser gain media.
An additional, non-invasively pumped crystal and waveguide was fabricated as a control.
2D cartoons of these two different ``lollipop'' geometries are shown in Fig.~\ref{fig:cross_section}.
Both waveguides were home-manufactured from 13-cm lengths of fused-quartz rod, 5 mm in diameter, using traditional lens-making techniques\cite{twyman88}.
On the input end, both waveguides have a polished flat face (oriented at right angles to the rod's axis).
The butt-coupled waveguide has an identical face at its other, outputting end,
whilst the outputting end of the ``spear-headed'' waveguide comprises three teardrop-shaped flat facets.
The waveguide's length as well as the layout and wiring of the LED's metal-core PCB, help reduce the maser crystal's exposure to transient magnetic flux generated by the electrical current flowing through the LED.
The use of magnetic shielding (mumetal) could have permitted a far smaller separation, but was not implemented.

\begin{figure}[hb]
    \centering
    \includegraphics[width=1\linewidth]{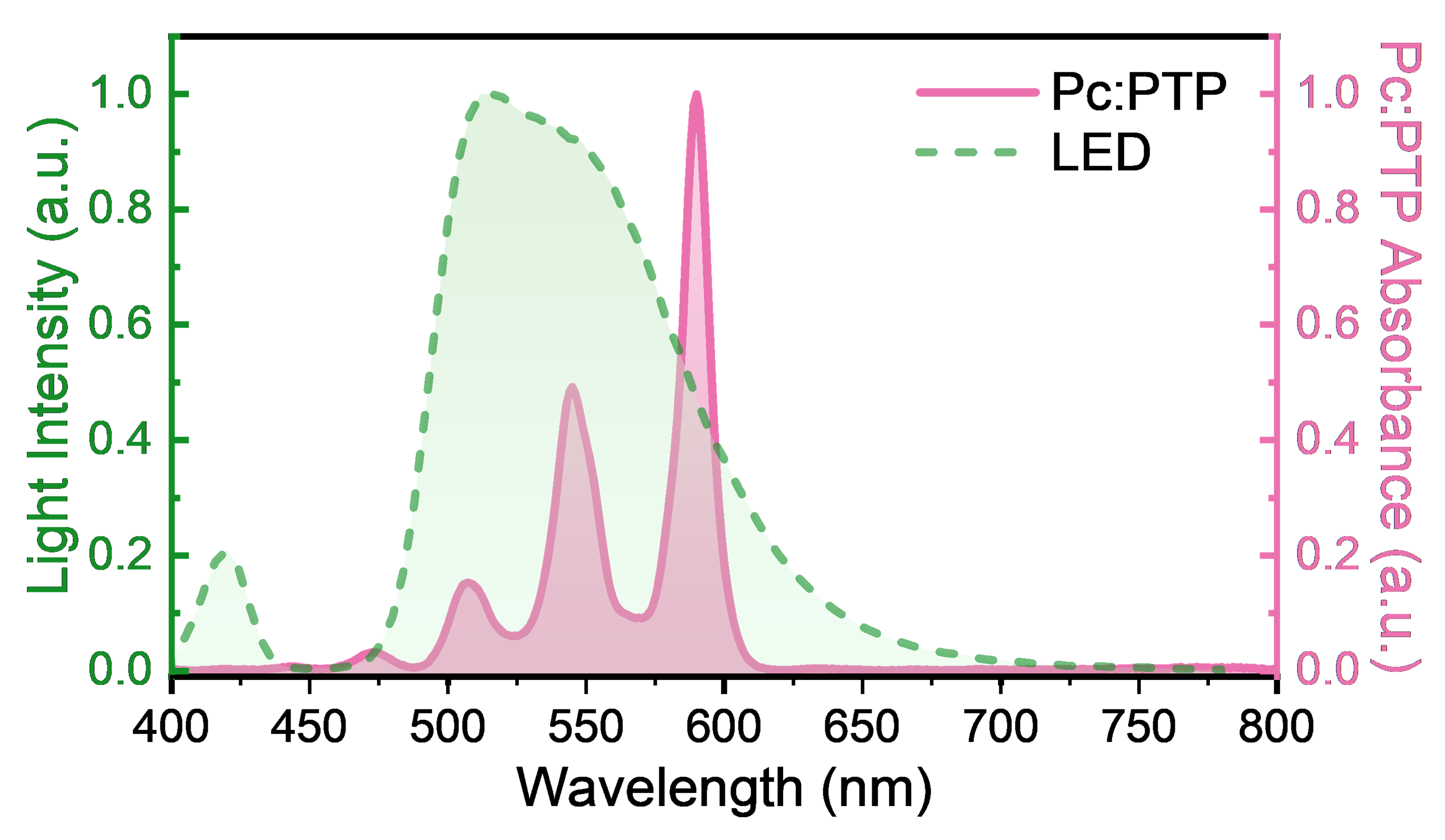}
    \caption{Output spectrum of LED (from manufacturer's datasheet\cite{osmar}) compared against the absorption spectrum of ptc:ptp\cite{supamas_2020}.}
    \label{fig:LED_crystal_spectrum}
\end{figure}

\begin{figure*}[ht!]
    \centering
    \hfill
    \includegraphics[width=1\textwidth]{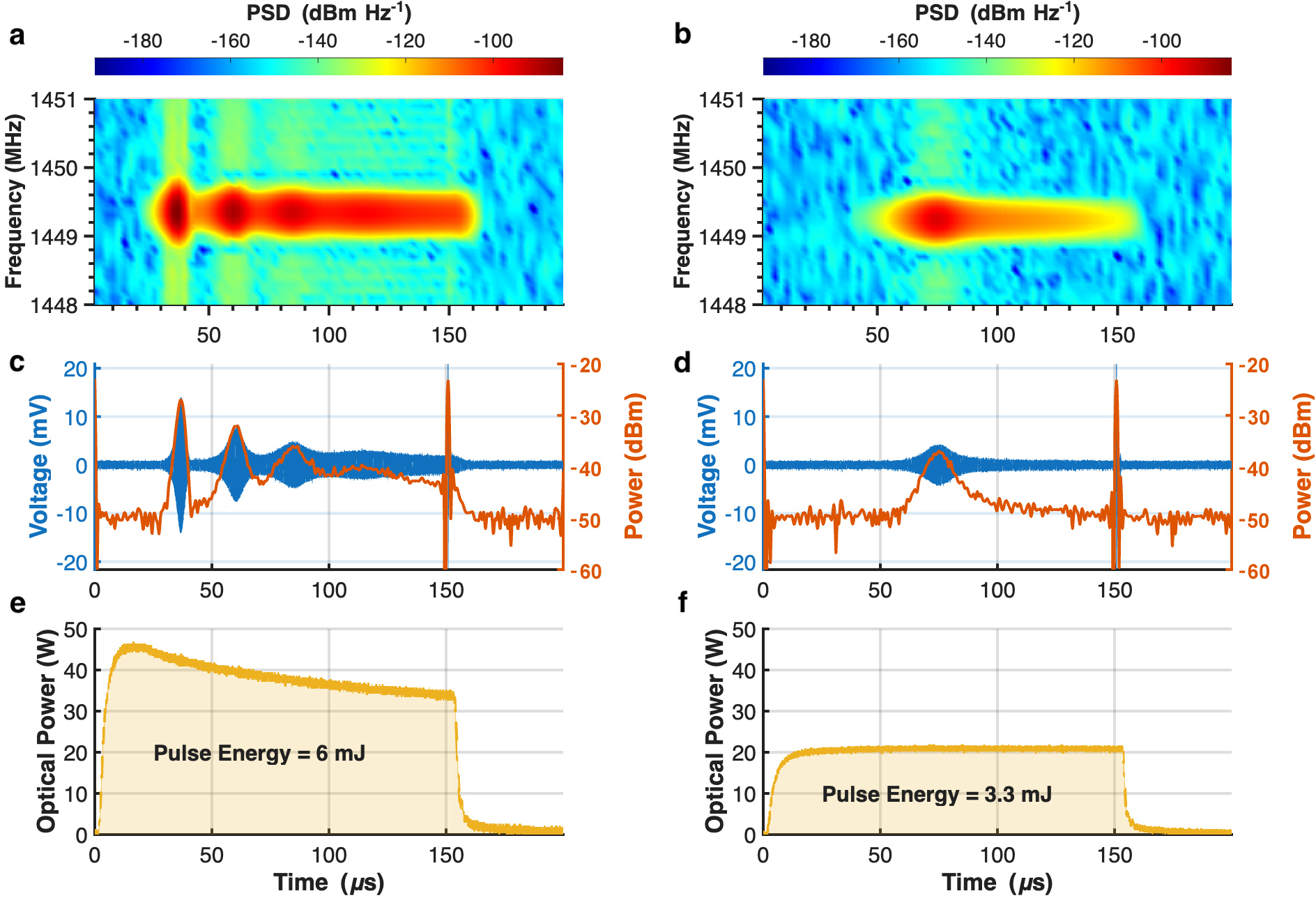}
\caption{Pulsed oscillations from the ``spear-head'' invasively pumped ptc:ptp maser under two different pump powers/energies. The left column (\textbf{a, c, e}) corresponds to a LED pump energy (distributed over 150 $\mu$s) of 6~mJ; the right column (\textbf{b, d, f}) corresponds to a LED pump energy (distributed over the same duration ) of $3.3$~mJ; the latter energy lies just above maser threshold. The top row (\textbf{a, b}) shows the output's power spectral density (PSD, in dBm~Hz$^{-1}$), derived from the short-time-Fourier transform of the outputted signal, taken using Hamming windows containing $2.5\times10^4$ samples with 50~\% overlap and $5\times10^4$ DFT points, generated in MATLAB R2025B. The middle row (\textbf{c, d}) shows the voltage output measured directly by a Keysight DSOX 6002A oscilloscope in blue, and the corresponding RMS power envelope in red, taken as an envelope over windows of 3000 samples. The lower row (\textbf{e, f}) shows the instantaneous optical pump power supplied by the LED as a function of time; the area-under-the-curve of each profile, equal to the total optical energy supplied by the LED, is stated. }
\label{fig:masing}
\end{figure*}

A ptc:ptp crystal was grown atop the outputting end of the spear-headed quartz waveguide by the vertical Bridgman method, as has been described in detail in previous work~\cite{Wu2020InvasiveCoupling}.
This method suspends the waveguide inside the glass growth vial in which the ptc:ptp powder is held, allowing a crystal to grow onto and around it,
the glass waveguide forming part of the containing mould against which the crystal grows.
An equivalent crystal with a butt-coupled waveguide was grown from the same stock mix of ptc and ptp powder using the same equipment and protocols (Bridgman oven temperature and drop speed). 
The optical source chosen to pump the grown ptc:ptp maser crystal is an InGaN-based phosphor-conversion LED chip,
\textit{viz.}~an OSRAM LE CG P2AQ)~\cite{osmar}, producing high-luminance yellow-green 565-575~nm light, intended for use in video projectors/displays.
Its output spectrum, relative to the absorption spectrum of ptc:ptp\cite{nelson1981laser,Ng2023MoveElectrodynamics},
is shown in  Fig.~\ref{fig:LED_crystal_spectrum}.
The LED's outputting surface is rectangular with dimensions of 3.2~mm $\times$ 2.6~mm,
which is fully covered by the flat inputting end face of each 5-mm-in-diameter quartz waveguide.
`Santovac 5' polyphenyl ether diffusion-pump oil is used as an inert optical coupling between the LED and waveguide. 
Further details on how the LED is driven and its output quantified are provided in Appendices D and E. 

The microwave resonator used in this work, as shown in Fig.~\ref{fig:anatomy}, closely resembles that used in ref.~\onlinecite{Wu2020Room-TemperatureConcentrator} though with a transparent fused quartz waveguide reminiscent of ref.~\onlinecite{Wu2020InvasiveCoupling}.
Made of OFHC copper, its cylindrical outer cavity serves to confine the TE$_{01\delta}$ mode of a SrTiO$_3$ (= ``STO'') dielectric ring supported concentrically on a post (made of cross-linked polystyrene) inside of it. This ring has the following dimensions: height 8.2~mm, OD 12.6~mm, ID 6.0~mm, resulting in a mode volume approximately twice that of the ring (whose ID was 4.0~mm) used
in ref.~\onlinecite{Wu2020Room-TemperatureConcentrator}.
The prepared ptc:ptp crystal and waveguide enter the cavity from below through a concentric hole in the cavity's base plate.
The mode's frequency can be adjusted by raising/lowering the height of the cavity's ceiling, and is tuned to the 1.4496 GHz
frequency of ptc:ptp's  $\text{T}_\text{X}-\text{T}_\text{Z}$ transition.
Two inductive-coupling loop probes (formed on the ends of RG402 semi-rigid coax) are inserted into the cavity.
The reflections (``return loss'') off each probe, $S_{11}$ and $S_{22}$, and the transmission between them through the cavity, $S_{21}$, are measured using an HP 8753A vector network analyzer (VNA).
Doing so allows the mode's frequency and loaded quality factor, $Q_\text{L}$,
to be measured and thereupon its intrinsic quality factor, $Q_\text{0}$, to be inferred. 
The diameter, orientation and insertion depth of each loop were adjusted such that $S_{11} \approx 0.05$
and $S_{22} \approx 0.3$ for the ``driving'' and ``listening'' loop probes, respectively. 
For both the spear-tip and butt-couple geometry, we found that $Q_\text{L} = 6000 \pm 200$.
[Because STO's permittivity is so high, the difference in how these two geometries load, dielectrically, the bore of the STO ring is minimal.]
To avoid interference (seeding of the maser oscillation), the VNA is always disconnected (using high-isolation mechanical coaxial switches) from the cavity before the LED is fired.

The voltage signal generated by the cavity's listening loop probe is either (i) directly recorded by a Keysight DSOX 6002A oscilloscope (sampling at 5~GSa~s$^{-1}$) --as shown in Fig.~\ref{fig:masing}, or 
(ii) amplified then downshifted using a mixer to 70~MHz and band-pass filtered (bandwidth = 4 MHz;
the signal processing here equates to a single-stage superheterodyne receiver),
then recorded on a RIGOL DS1104 100-MHz oscillscope --as shown in Fig.~\ref{fig:sim-exp_waterfal}.
Using the above-described set-up, maser oscillations of tuneable duration and intensity could be observed on the oscilloscope.
When the LED is set to output 6~mJ pulses, 150~$\mu$s in duration, the strongest maser oscillations are recorded.
An example (single shot) is presented in Fig.~\ref{fig:masing}(a) and (c).
In this case, the maser pulse starts approximately 30~$\mu$s after the LED is provided with current (at $t =0$), which is significantly longer than the few $\mu$s start up times reported for other ptc:ptp maser systems~\cite{Salvadori2017NanosecondMASER,Ng2023MoveElectrodynamics}.
Looking at Fig.~\ref{fig:masing}(e), we can at least partially attribute this long start-up time 
to the slow turn on of the LED's drive current and thus optical output (taking around $10~\mu$s).
We in turn attribute this slow turn on to parasitic inductance in the cables running between the LED
and its current-drive box.[Better routing (twisting) of these cables could have reduced the inductance but
was not implemented.] 
Here, the maser oscillation displayed Fig.~\ref{fig:masing}(a) and (c) also exhibits pulsed amplitude modulation
akin to the Rabi oscillations exhibited by similar ptc:ptp-loaded STO-ring resonators
when pumped (more powerfully and abruptly) by Q-switched laser beams\cite{Breeze2017Room-temperatureStates}.
The amplitude modulation smears out within the 150~$\mu$s duration of the applied pump,
leaving continuous-wave maser oscillation\cite{Wu2020Room-TemperatureConcentrator} at relatively a low amplitude and power until the optical pump is terminated.   
This quasi-continuous wave (quasi-CW) operation has been previously reported in two other ptc:ptp maser systems~\cite{Wu2020Room-TemperatureConcentrator,Sathian2025LED-pumpedMaser} and indicates that the LED  provides sufficient optical pump power/luminance to sustain maser oscillation even as the $\text{T}_\text{X}:\text{T}_\text{Z}$ population ratio decreases due to bottlenecking~\cite{Wu2019UnravelingField}. The maser output was found to be stable over hours of pulsed operation (at a repetition rate of 1 Hz), provided the cavity is re-tuned (manual adjustment of ceiling height) from time to time to compensate for the STO dielectric slowly warming~\cite{Wu2020Room-TemperatureConcentrator}. A more thermally stable dielectric, an all-metallic resonator~\cite{xiang2025all}, or some basic form of forced-air cooling/temperature control, would be expected to provide truly uniterrupted, autonomous operation. The transient peaks observed at the start and end of the maser burst are attributed to electromagnetic interference (inductive cross-talk) from the LED's current supply (driver) lines;
these transients have no detectable effect on the optical output of the LED [Fig.~\ref{fig:masing}(e)].

Reducing the optical output power to 3.3~mJ over the same pulse duration results in a much weaker (``runt'') maser burst, which starts later at approximately 60~$\mu$s after at $t =0$), and decays more rapidly. The lengthened turn-on time, lack of Rabi-like amplitude modulation, reduced amplitude, and lack of sustained, quasi-continuous masing all concur with the literature~\cite{Salvadori2017NanosecondMASER, Breeze2017Room-temperatureStates}. This level of pump energy/power lies just above the threshold for masing: dialling down the voltage on the capacitor (whilst keeping the same pulse duration) causes the runt to disappear completely.
When the same resonator is loaded with a cylindrical ptc:ptp crystal butt-coupled to a cylindrical waveguide,
as per the right side of Fig.~\ref{fig:cross_section},
no maser oscillations (not even runts) are observed, even when the pulse energy is increased to 6~mJ (at 150~$\mu$s duration).
Exceeding this pulse energy and/or duration risks blowing up the LED (known from direct experience.)

At the relatively low levels of instantaneous power/intensity provided by LEDs (in comparison to Xe flash lamps or pulsed/Q-switched lasers),
ptc:ptp does not bleach\cite{Takeda2002Zero-fieldCrossing}. Its gain and thus the maser's cooperativity are expected to grow linearly with applied optical pump power~\cite{Oxborrow2012MaserAssembly}. Accepting this linearity, the most intense maser bursts for our
spear-head geometry (Fig.~\ref{fig:anatomy}, the left side of Fig.~\ref{fig:cross_section}),
recorded at 6~mJ per pulse, should have a cooperativity of $\sim$2. For an intrinsic cavity $Q_0$ of $6,000$, this corresponds to a magnetic $Q_\text{m}$ of $3,000$~\cite{A.E.Siegman1964MicrowaveMasers}. For comparison, ref.~\onlinecite{Breeze2015EnhancedMasers} reports a ptc:ptp maser, pumped by a xenon flash lamp producing a peak useable optical output power of 60-70~W --though only $\sim$40~W of this gets absorbed; it exhibits a $Q_\text{m}$ of $4,500$ in a resonator with a $Q_\text{L}$ of 8,900, so attaining a comparable cooperativity to our LED-pumped spear-head geometry.

\section{Theory (Optimization of Cooperativity)}

The theory of optically-pumped masers is here reformulated towards justifying our design choices and inferences (from experimental data/observations). The power dissipated from an electromagnetic mode, at frequency $f_\textrm{mode} = \omega / (2 \pi)$, within a cavity is
\begin{equation}
P_\textrm{dis.} = \frac{U}{Q_\textrm{L}} \omega = \frac{ \frac{1}{2} \mu_0 \iiint_\textrm{cav.} |\textbf{H}(\vec{\textbf{r}})|^2 \textrm{d}V} {Q_\textrm{L}} 2 \pi f_\textrm{mode},
\label{eqn:power_dis}
\end{equation}
where $U$ is the energy stored in the cavity,
$Q_\textrm{L}$ is the mode's loaded quality factor,
and $\textbf{H}(\vec{\textbf{r}})$ is the (assumed to be linearly polarised) magnetic field strength (a vector field, dependent on position $\vec{\textbf{r}}$) within it.
Here, for simplicity, we assume that the dielectric medium/media filling
the inside of the cavity (but for the paramagnetic maser crystal itself) is non-magnetic. 
Using the un-numbered equation above equation (5-5-26) in ref.~\onlinecite{A.E.Siegman1964MicrowaveMasers}
[at zero detuning, \textit{i.e.}~``on resonance'',
for an allowed spin $S = 1$ transition whose magnetic dipole moment
is everywhere aligned with (\textit{i.e.}~parallel to) $\textbf{H}$],
the power deposited into the same microwave mode by stimulated emission is given by
\begin{equation}
P_\textrm{gen.} =
\frac{1}{4} T_2 \gamma^2 (h f_\textrm{mode}) \iiint_\textrm{cav.} \Delta n(\vec{\textbf{r}}) \; |\textbf{H}(\vec{\textbf{r}})|^2 \textrm{d}V,
\label{eqn:power_gen_siegman}
\end{equation}
where $T_2$, in Siegman's parlance, is the spin-spin relaxation time, $\gamma$ is the electronic gyromagnetic ratio, and $\Delta n(\vec{\textbf{r}}) $ is the spin-polarization (otherwise known as the population inversion) density as a function of position $\vec{\textbf{r}}$.
In general, both  $\Delta n(\vec{\textbf{r}})$ and $\textbf{H}(\vec{\textbf{r}})$ will vary with position $\vec{\textbf{r}}$ throughout the maser's gain medium (henceforth referred to, more succinctly, as the ``crystal'' or ``x-tal''). Since $\Delta n(\vec{\textbf{r}})$ is zero everywhere outside of the crystal, the volumetric integral need not extend beyond it. To take into the account the effects of inhomogeneous broadening and/or spin diffusion (reducing the number of pumped, spin-polarized molecules that can contribute to stimulated emission at any one moment), $T_2$  is often replaced by $T_2^*$; we henceforth adopt this replacement without any deeper justification.

The maser's cooperativity, defined as the rate of stimulated emission divided by the rate of dissipation, is thus 
\begin{equation}
C = \frac{P_\textrm{gen.}}{P_\textrm{dis.}} = \frac{\gamma^2 h}{4 \pi \mu_0} T_2^* Q_\textrm{L}\frac{\iiint_\textrm{x-tal} \Delta n(\vec{\textbf{r}}) \; |\textbf{H}(\vec{\textbf{r}})|^2 \textrm{d}V}{\iiint_\textrm{cav.} |\textbf{H}(\vec{\textbf{r}})|^2 \textrm{d}V} .
\label{eqn:coop1}
\end{equation}
The first quotient in the expression on the right-hand side contains immutable physical constants;
it shall henceforth be denoted as $[.]$ to save space. 
In the special case of a uniformly spin-polarized maser crystal, the quotient on the far right formed by the above expression's two volume integrals  reverts to just $\Delta n \, \zeta$, where $\zeta$ is the so-called magnetic filling factor of the maser crystal, and equation \ref{eqn:coop1} reverts (taking $C = 1$) to the more familiar mean-field expressions for the maser-threshold condition found in the literature (to within a factor of 2 at least).\cite{Oxborrow2012MaserAssembly,Sherman2021PerformanceMasers,Arroo2021PerspectiveMasers}

In the case of a non-bleaching ``linearly pumpable'' medium (with respect to its photo-excitation and subsequent spin dynamics), analogous to a 4-level laser system, one would expect the local spin polarization density $\Delta n$ to scale in proportion to the optical excitation (= absorption) rate, namely that 
$\Delta n(\vec{\textbf{r}}) = p_\textrm{abs.}(\vec{\textbf{r}}) \; \Theta \;T_1^\textrm{eff.} / (h f_\textrm{opt.})$,
where $p_\textrm{abs.}(\vec{\textbf{r}})$ is the local density of absorbed optical power [in units of watts per unit volume, as a function of position $(\vec{\textbf{r}})$],
$h f_\textrm{opt.} = h c / \lambda_\textrm{opt.}$ is the energy of a pump photon;
$\Theta$ equals the probability, per pump photon absorbed, of a maser molecular arriving at the upper maser level minus the probability of it arriving at the lower maser level; for a pentacene-doped \textit{para}-terphenyl maser operating on its X-Z transition,  
this equals the intersystem crossing (ISC) yield multiplied by the difference between the
initial fractional populations of the X and Z levels of pentacene's triplet ground state;
and, finally, $T_1^\textrm{eff.}$ is the effective spin-lattice relaxation time,
quantifying the longevity of the pentacene molecule's optically excited (longitudinal) spin polarization.
An explicit expression for $T_1^\textrm{eff.}$, in terms of the spin-lattice rates between and lifetimes of the sublevels of pentacene's triplet ground state, is provided in ref.~\onlinecite{Oxborrow2012MaserAssembly}.

It is insightful to replace
$p_\textrm{abs.}(\vec{\textbf{r}})$ by $P_\textrm{abs.} \times \bar{p}_\textrm{abs.}(\vec{\textbf{r}})$, where $P_\textrm{abs.}$ is the total absolute optical power absorbed by the whole maser crystal and $\bar{p}_\textrm{abs.}(\vec{\textbf{r}})$, with units of inverse volume, is the normalised spatial distribution ($\iiint \bar{p}_\textrm{abs.}(\vec{\textbf{r}}) \textrm{d} V = 1$) describing how  this total absorbed power is distributed over space.
Also, the volume integral in the denominator of the r.h.s.~of equation \ref{eqn:coop1} above can be re-expressed as
$V_\textrm{mode} \times |\textbf{H}|^2_\textrm{max}$, where $V_\textrm{mode}$ is (by definition) the maser mode's magnetic mode volume and $|\textbf{H}|^2_\textrm{max}$ is the maximum value of $|\textbf{H}(\vec{\textbf{r}})|^2$ (evaluated at the mode's magnetic antinode). 
These manoeuvres enable equation~\ref{eqn:coop1} to be recast as
\begin{equation}
C  = [.] \frac{T_1 T_2^*  \Theta }{f_\textrm{opt.}}
\frac{Q_\textrm{L} }{V_\textrm{mode} } P_\textrm{abs.}
\iiint_\textrm{x-tal}
\bar{p}_\textrm{abs.}(\vec{\textbf{r}}) \; \frac{|\textbf{H}(\vec{\textbf{r}})|^2} {|\textbf{H}|^2_\textrm{max}} \textrm{d} V. 
\label{eqn:coop2}
\end{equation}
The first explicit quotient on the r.h.s.~provides a figure of merit on the maser's ``chemistry'':
it contains parameters that control the photo- and spin-dynamics of the chosen maser gain medium.
So far as they can be modified/engineered, one strives to maximize all of the factors in the above expression.  
The volume integral therein is, in words, the fractional magnetic strength squared averaged over the spatial distribution
of the absorbed optical power; we shall henceforth refer to this important dimensional quantity as the ``\textbf{coincidence}''
(between the maser mode's a.c.~magnetic field and where the gain medium gets optically pumped) and denote it by 
$\Gamma \equiv \iiint_\textrm{x-tal}
\bar{p}_\textrm{abs.}(\vec{\textbf{r}}) \; |\textbf{H}(\vec{\textbf{r}})|^2 / |\textbf{H}|^2_\textrm{max}  \textrm{d} V$.
Note that, by construction, $\Gamma \leq 1$.
For a given gain medium, design of microwave cavity, and applied/absorbed optical power, 
we see that, provided the latter is not so strong as to exhaust (bleach) or overheat the gain medium,
the maser's optical engineering (toward maximizing the cooperativity)
boils down to maximizing $ P_\textrm{abs.} \times \Gamma$. We now consider two limiting cases:

\textit{Weak pump limit:} When the available optical pump power is small, we see that the greatest  cooperativity is achieved by maximizing the so-called magnetic Purcell factor\cite{Breeze2015EnhancedMasers} defined as $Q_\textrm{L} / V_\textrm{mode} $,
whilst positioning the small amount of maser crystal that can be pumped to where $ |\textbf{H}(\vec{\textbf{r}})|^2 $ is greatest,  
whilst focussing/guiding all of the available pump light towards this point, such that $\Gamma \simeq 1$. 

\textit{Unlimited pump:} 
When greater optical pump power is available, the optical absorption must be spread out to avoid
exhausting (bleaching) or overheating the gain medium.
If the maser operates in burst/pulsed mode, with little movement/diffusion of the absorbed optical energy within the pump pulse's duration, the temperature attained at any point in the maser crystal will rise monotonically with the absorbed energy density. Towards maximizing the cooperativity, it then becomes advantageous for (i) the maser crystal to extend throughout the internal volume of the cavity (excluding those regions already occupied by the dielectric) wherever the fraction $ |\textbf{H}(\vec{\textbf{r}})|^2 / |\textbf{H}|^2_\textrm{max} $ is substantially greater than zero, and for (ii) this extended crystal to be illuminated uniformly to just below the maximum absorbed power density that thermal/photochemical limits will allow. 
Here, it is immediately acknowledged that achieving such uniformity of illumination
is a non-trivial optical-engineering challenge. 
Within this scenario, one should expect that 
$P_\textrm{abs.}^\textrm{max} = \alpha_\textrm{x-tal} \, V_\textrm{illum.}$,
where $\alpha_\textrm{x-tal}$ depends on the maser crystal's physical and chemical properties
(such as its volumetric heat capacity and melting/charring temperature),
and $V_\textrm{illum.}$ is the illuminated volume over which the pump light gets absorbed. 
With this substitution introduced, we arrive at
\begin{equation}
C  = [.] \frac{T_1 T_2^* \Theta \alpha_\textrm{x-tal} }{f_\textrm{opt.}}
Q_\textrm{L} P_\textrm{abs.}
\iiint_\textrm{illum.}
\frac{|\textbf{H}(\vec{\textbf{r}})|^2} {|\textbf{H}|^2_\textrm{max}}  \textrm{d}V  / V_\textrm{mode} . 
\label{eqn:coop3}
\end{equation}
where now the volume integral extends over only those parts of the crystal that are (uniformly) illuminated.
The factor  $\eta_\text{fill.} \equiv \iiint_\textrm{illum.}
|\textbf{H}(\vec{\textbf{r}})|^2 /|\textbf{H}|^2_\textrm{max}  \textrm{d}V  / V_\textrm{mode}$ serves
as the maser crystal's effective magnetic filling factor. 
Within this scenario of unlimited optical pump power,
optimizing the cooperativity involves maximizing the $Q_\textrm{L} \times \eta_\text{fill.}$ product,
analogous to maximizing the gain of a tradition solid-state maser amplifier \cite{A.E.Siegman1964MicrowaveMasers},
or the sensitivity of an EPR spectrometer\cite{Poole1983} (for an unlimited
volume of available sample). 

\section{Simulation and Analysis}

\subsection{RF and Optical}

The TE$_{01\delta}$ mode's a.c.~magnetic field, as is largely defined by the shape of the cavity's STO ring
(and the location of the cavity's interior copper walls relative to it) are calculated using a FEM-based solver of Maxwell's equations, namely COMSOL Multiphysics.
These simulations provide the vector field
$\vec{\textbf{B}}(\vec{\textbf{r}}) = \mu_0 \vec{\textbf{H}}(\vec{\textbf{r}}) $ and
its modulus squared (proportional to the magnetic energy density) through the cavity's interior.
By convention, the field is normalised such that 
$\iiint_\textrm{cav.} \frac{1}{2} \mu_0 |\textbf{H}|^2 \textrm{d}V =  \frac{1}{2} (|\textbf{B}|^2 / \mu_0) \textrm{d}V = 1$~J. 
See Appendix~A for further details.

\begin{figure}[h]
    \hfill
    \includegraphics[width=\linewidth]{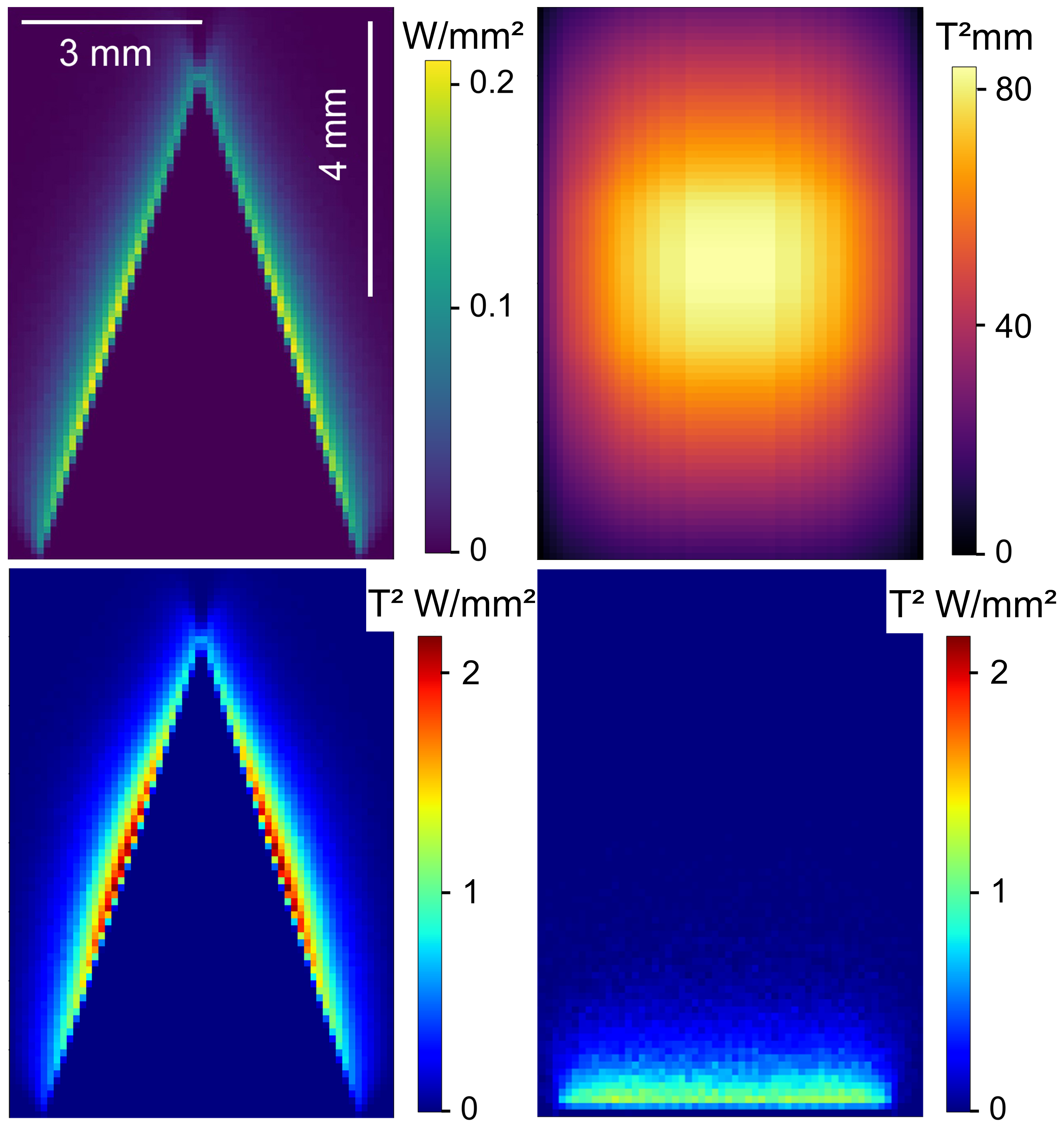}
    \caption{
    Top left: Simulation of the normalised projected absorbed optical power density,
    $p_{\mathrm{abs.},2\mathrm{D}}$, in the invasively pumped ptc:ptp sample,
    using PVtrace.
    Top right: COMSOL simulation of the normalised projected magnetic flux density,
    $|\vec{\mathbf{B}}|^{2}_{2\mathrm{D}}$,
    inside the bore of the STO dielectric resonator.
    The projected quantities are obtained by integrating the corresponding 3D distributions
    along the viewing direction into the page,
    $p_{\mathrm{abs.},2\mathrm{D}}
    =\int p_{\mathrm{abs.}}\,dl$
    and
    $|\mathbf{B}_{2\mathrm{D}}|^{2}
    =\int |\mathbf{B}|^{2}\,dl$.
    The displayed projected densities are normalised such that
    $\int p_{\mathrm{abs.},2\mathrm D}\,dA=1~\mathrm W$
    and
    $\frac{1}{2\mu_0}\int |\vec{\mathbf B}|^2_{2\mathrm D}\,dA
    =1~\mathrm J$.
    Bottom panels:
    Comparison of the projected three-dimensional overlap,
    $\int p_{\mathrm{abs.}}
    |\vec{\mathbf{B}}|^{2}\,\mathrm{d}l$,
    for invasive pumping (left) and butt-coupling (right).
    }
    \label{fig:invasive_vs_butt}
\end{figure}

To evaluate and understand the coincidence integral constructed in the previous section, we simulate the spatial distribution of the optical energy density absorbed by the ptc:ptp crystal, $p_\text{abs.}(\vec{\textbf{r}})$ (in units of W/mm$^3$), for both the invasive spear-head and butt-coupled configurations, using PVtrace open-source ray-tracing software~\cite{Farrell2008,FarrellPVTraceOriginal,verma2023ray,PVTraceModified}. 
Within our simulations, we treat the LED as a flat rectangular lambertian light source, emitting through a layer of optical coupling liquid into the waveguide. Light from the LED that is successfully captured by the waveguide then propagates along it by total internal reflection until it meets the waveguide's terminating ptc:ptp crystal.  
In our initial simulations here, we assume that 0.1\%~ptc:ptp is a linear (bleaching effects are ignored) optical absorber with a Beer-Lambert absorption coefficient of 2~mm$^{-1}$, independent of wavelength, corresponding
to a molecular cross-section of $0.7 \times 10^{-17}$~cm$^2$. [This value is approx.~3 times smaller than what
ref.~\onlinecite{nelson1981laser} measured at 532~nm, but is what we measured\cite{ZikeChengMSc} using a UV-vis spectrophotometer on samples taken from the same growth vial in which the spear-head lollipop grew.] 
Neither the birefringence of \textit{para}-terphenyl nor the anisotropy (both direction and polarisation dependence) of pentacene's optical absorption (when dissolved as a substitutional dopant in the former) are explicitly included.  
To more easily visualize where the optical pump gets absorbed, we here first simulate a simplified, two-faced wedge geometry.
The two facets of this double wedge have the same angle of slope (relative to the cylindrical waveguide optical axis) as the three facets of the experimental spear head.
This choice allows for clear and intuitive projections of the absorbed optical energy density
onto the 2D page without hugely affecting the model's fidelity.

Simulated projected distributions of the absorbed optical power density, \(p_{\mathrm{abs.}}(\vec{\textbf{r}})\),
the magnetic flux density, \(|\vec{\mathbf{B}}(\vec{\textbf{r}})|^2\),
and the projected distribution of their product are shown in Fig.~\ref{fig:invasive_vs_butt}. 
Here, each of the original densities, a scalar field over 3D space, is projected onto the 2D page by integrating along 
the ``depth'' direction (normal to the page) through the resonator:
\(p_{\textrm{abs.},\mathrm{2D}} \equiv \int {p}_{\mathrm{abs.}}\,\mathrm{d}l\), and \(|\vec{\mathbf{B}}|^2_{\mathrm{2D}} \equiv \int|\vec{\mathbf{B}}|^{2}\,\mathrm{d}l\), where \(l\) ranges over the depth direction. The two lower plots in Fig.~\ref{fig:invasive_vs_butt} are obtained by projecting
the product of $p_{\textrm{abs.},\mathrm{2D}}$  and $|\vec{\mathbf{B}}|^2_{\mathrm{2D}}$ onto the 2D page,
\textit{i.e.},
$\int (p_{\mathrm{abs.}} \times|\vec{\mathbf{B}}|^{2}) \,\mathrm{d}l \equiv  [p_{\mathrm{abs.}} \times |\vec{\mathbf{B}}|^2]_{\mathrm{2D}}$,
rather than multiplying $p_{\textrm{abs.},\mathrm{2D}}$ and  
$|\vec{\mathbf{B}}|^2_{\mathrm{2D}}$ together.

Combining optical and RF simulations in this manner allows one to both map 
$[p_{\mathrm{abs.}} \times |\vec{\mathbf{B}}|^2]_{\mathrm{2D}}$
and to estimate the normalised utilisation or overlap factor defined as:
$\Delta = \iiint_{\textrm{x-tal}} p_{\mathrm{abs.}}(\vec{\textbf{r}})\, |\vec{\mathbf{B}}(\vec{\textbf{r}})|^{2}\,\mathrm{d}V
\Big/ \iiint_{\mathrm{cav.}}|\vec{\mathbf{B}}(\vec{\textbf{r}})|^{2}\,\mathrm{d}V$,
which provides a measure of the overall efficiency of the pumping geometry simulated. 
We find that, for 1 joule of a.c.~magnetic field energy,
$\Delta = 2.26 \times 10^{-3}$ T$^2$ for the butt-coupled configuration,
while, for the invasively pumped configuration, it evaluates to $\Delta = 4.60 \times 10^{-3}$ T$^2$,
amounting to a 2.08 times improvement --despite the large excluded volume
(where ptc:ptp crystal can no longer reside) from the presence of the invading waveguide.  
Given that the invasively pumped variant exhibits a cooperativity of $\sim$2,
this result implies that the butt-coupled variant should lie just
below the $C = 1$ threshold for maser oscillation, consistent with experiment. 
For reference, we find that, for completely uniform optical absorption throughout the crystal,
$\Delta = 4.56\times10^{-3}$ T$^2$. 
Our plots of $[p_{\mathrm{abs.}} \times |\vec{\mathbf{B}}|^2]_{\mathrm{2D}}$ in the
lower half of Fig.~\ref{fig:invasive_vs_butt} reveal that the ptc:ptp gain medium
is very poorly utilised for either shape of waveguide.
This suggests that there remains substantial headroom for improvement through, 
for example, the design and realization of more complicated invasive optics (or ``injectors'')
capable  of supplying light to a larger volume of crystal. 

\subsection{Polychromatic ray-tracing simulation and Monte-Carlo evaluation of cooperativity}

A more sophisticated yet in many ways more universal and simpler ray-tracing simulation is here developed (still using PVtrace).
This includes both the LED's true polychromatic output spectrum and the wavelength dependence of ptc:ptp's absorption,
as are both shown in Fig.~\ref{fig:LED_crystal_spectrum}. Please see Appendix B for a detailed description on how PVtrace is configured.
\begin{figure}[ht!]
    \centering
    \hfill
    \includegraphics[width=1.0\linewidth]{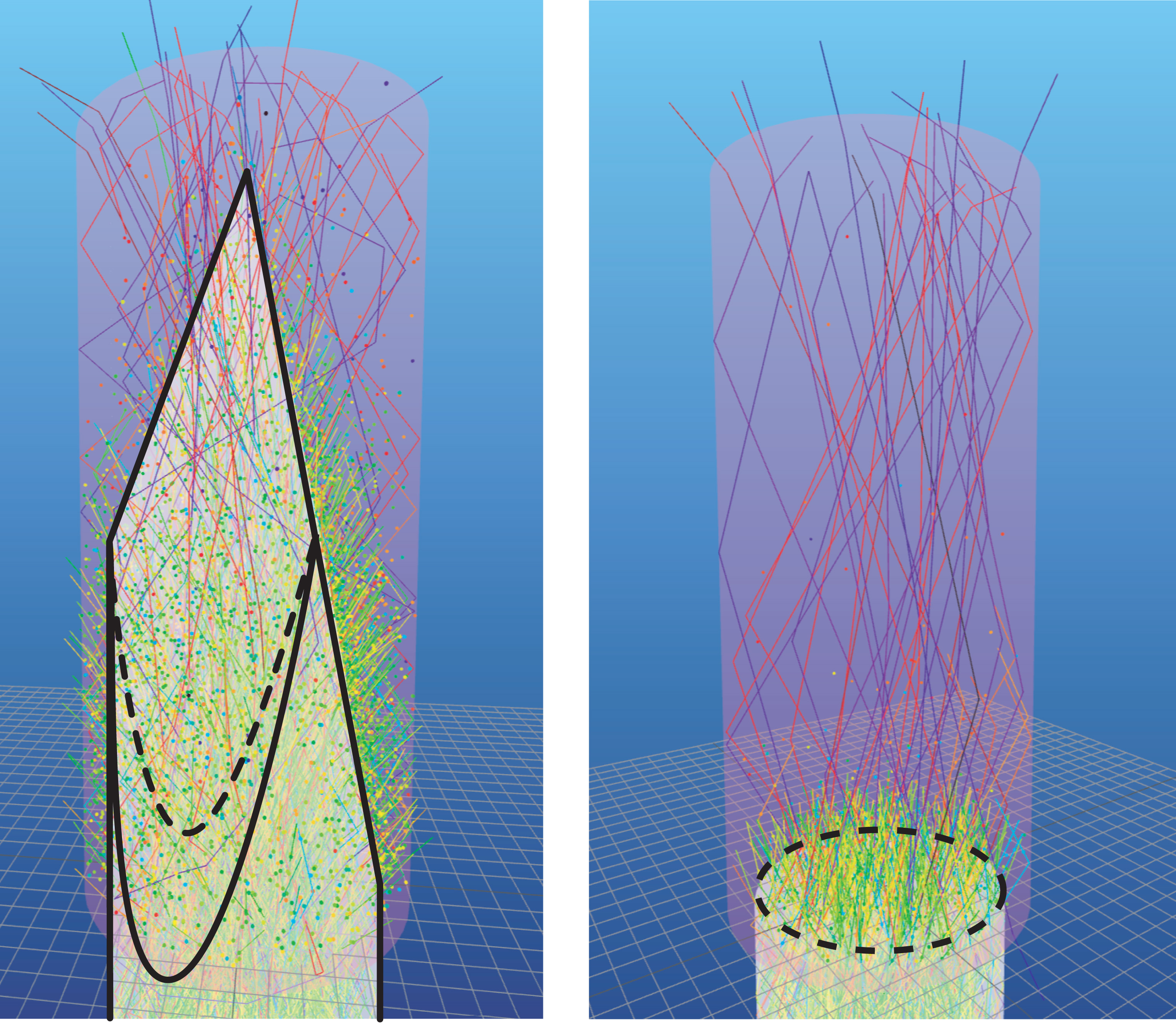}
\caption{Polychromatic ray-tracing simulations of optical propagation and absorption. Left: Invasive spear head. Right: Butt-coupled. The orientation of the former is chosen to make clear the ``fringe'' of rays emitted from each of the waveguide's facets and absorbed by the ptc:ptp crystal. }
\label{fig:sim-3D_ray-trace}
\end{figure}
Experimentally, the optical power emitted by the pump source, $P_\textrm{emit.}$,
is far easier to measure and control than the fraction thereof, $P_\textrm{abs.}$,
that gets absorbed by the maser crystal. 
Provided the propagation of light from the pump source to the maser crystal
and its absorption therein are linear processes, one can formally re-express the absorbed optical power density as 
$p_\textrm{abs.}(\vec{\textbf{r}}) = P_\textrm{emit.} \alpha_\textrm{coupl.}(\vec{\textbf{r}})$,
where the ``coupling distribution'' $\alpha_\textrm{coupl.}(\vec{\textbf{r}})$, with units of inverse volume,
is equivalent to the normalised absorption distribution $\bar{p}_\text{abs.}(\vec{\textbf{r}})$ above, 
but goes further by taking into account that not all of the photons (light rays) emitted by the pump source will
make it to the maser crystal and get absorbed there.
Some will miss, some will propagate straight through without being absorbed;
thus, in general, $\iiint \alpha_\textrm{coupl.}(\vec{\textbf{r}}) \leq 1$.
This re-expression allows the cooperativity to be recast as:
\begin{equation}
C  = [.] \frac{T_1 T_2^*  \Theta }{f_\textrm{opt.}}
\frac{Q_\textrm{L} }{V_\textrm{mode} } P_\textrm{emit.}
\iiint_\textrm{x-tal}
\alpha_\textrm{coupl.}(\vec{\textbf{r}}) \; \frac{|\textbf{H}(\vec{\textbf{r}})|^2} {|\textbf{H}|^2_\textrm{max}}  \textrm{d}V. 
\label{eqn:strike}
\end{equation}
The integral 
$S \equiv \iiint_\textrm{x-tal}
\alpha_\textrm{coupl.}(\vec{\textbf{r}}) |\textbf{H}(\vec{\textbf{r}})|^2 / |\textbf{H}|^2_\textrm{max} \textrm{d}V $
on the right hand side of this equation is dimensionless and quantifies how successfully
the maser's optics transfers pump light from its source to where it can be most advantageously absorbed.
We call this dimensionless integral the ``\bf{strike fraction}\rm''. 

An optical ray-tracing simulator such as PVtrace generates a large, representative (= unbiased) collection  of rays from emission point to absorption point (or else a ``miss''). The method of Monte Carlo integration can now be invoked to convert the strike fraction's volume integral into a sum over this collection of $N$ simulated rays:

\begin{equation}
S = \lim_{N \rightarrow \infty}
\frac{1}{N}
\sum_{N \,\textrm{rays}}
\frac{|\textbf{H}(\vec{\textbf{r}}_\textrm{term.})|^2} {|\textbf{H}|^2_\textrm{max}}.
\label{eqn:Monte_Carlo}
\end{equation}
Here, $\textbf{r}_\textrm{term.}$ is where a photon gets absorbed and its associated ray terminates;
if a ray misses or passes through the crystal, its contribution to the above sum is zero.
Provided many rays can be simulated, the above expression provides
a method for directly and accurately evaluating the strike faction.  
The optical engineering needed to maximize the cooperativity, $C$,
thereupon boils down to maximizing $ P_\textrm{emit.} \times S$.

Like before, raising $ P_\textrm{emit.}$, requires increasing the volume over which pump light gets absorbed. 
For an invasive  waveguide$\rightarrow$crystal geometry, the absorption of pump light will be highly uneven due to the nature (exponential decay) of Beer-Lambert absorption, with light arriving predominantly from one side/direction. 
We define the effective volume, $V_\textrm{abs.}^\textrm{eff.}$ over which pump light gets absorbed through the following ``voxelization'' method: The space occupied by the crystal is divided into cubic voxels,
$s =$~0.8~mm in side length. [This size is chosen to be smaller than the optical injector's features, yet large enough to suppress statistical noise.]    
From the set of 3D positions at which rays are terminated (or else escape), as is generated direct by our PVtrace simulation, we determine, for each voxel, the number of absorptions, integer $n$, that fall within the voxel and the corresponding discrete volumetric density, $p = n / s^3$.
A histogram of the number of voxels with an absorption density of $p$ is thereupon plotted over $p$. The spikey, left-most (small $p$) region of the histogram (corresponding to empty and almost empty voxels) is ignored/deleted, removing those columns corresponding to $p$ below 46 rays/mm$^3$ (or 0.46\% of the 10,000 input rays simulated). 
An exponential function of the form $a \; \textrm{exp}(-b \, n)$ is  then fitted to the undeleted section of histogram. Voxels with $p$ greater than or equal to the fitted exponential's ``half-life'' (in units of $p$), namely $\textrm{ln}(2/b)$, are deemed to lie inside the effective volume. We find that the value of the
$V_\textrm{abs.}^\textrm{eff.}$ calculated in this way fluctuates by a standard deviation of less than 5\% from $s = 0.5$ to $s = 1$~mm, demonstrating that our method is robust against the exact choice of voxel dimension~$s$.

\begin{table}[t]
  \centering
  \begin{tabular}{|c|c|c||c|}
    \hline
Geometry & Strike fraction & Effective absorption & Product \\
 &  $S$ & volume $V_\textrm{abs.}^\textrm{eff.}$ [mm$^3$] & $S$ $\times$ $V_\textrm{abs.}^\textrm{eff.}$ [mm$^3$] \\
\hline
butt-couple  & 0.242 & 10.1 & 2.44 \\
 at bottom    & & & \\
 \hline
 butt-couple  & 0.745 & 10.1 & 7.53 \\
 at equator   & & & \\
 \hline
 spear head   & 0.439 & 33.4 & 14.66 \\
\bottomrule    
    \hline
  \end{tabular}
  \caption{Optical pumping performance for different geometries.}
  \label{tab:strike_table}
\end{table}

Our improved 3D ray-tracing simulations, whose results are shown in Table~\ref{tab:strike_table},
covered three different geometries:
[i] a butt-couple (the right side of Fig.~\ref{fig:sim-3D_ray-trace}) located at the bottom of the STO ring's bore.
[ii] butt-couple located 4.5~mm (thus about half way) up the same bore,
[iii] a spear head (the left side of Fig.~\ref{fig:sim-3D_ray-trace}), whose tear-drop facets start
2.7~mm below the bottom of the STO ring's bore.
We see from the table that the spear-head geometry exhibits a three times larger effective optical absorption
volume whilst providing a decent strike fraction. This quantitatively explains its lower
pump threshold and imperviousness to damage (from pump light) compared to the two butt-couple designs.

\begin{figure}[ht!]
    \centering
    \hfill
    \includegraphics[width=1.0\linewidth]{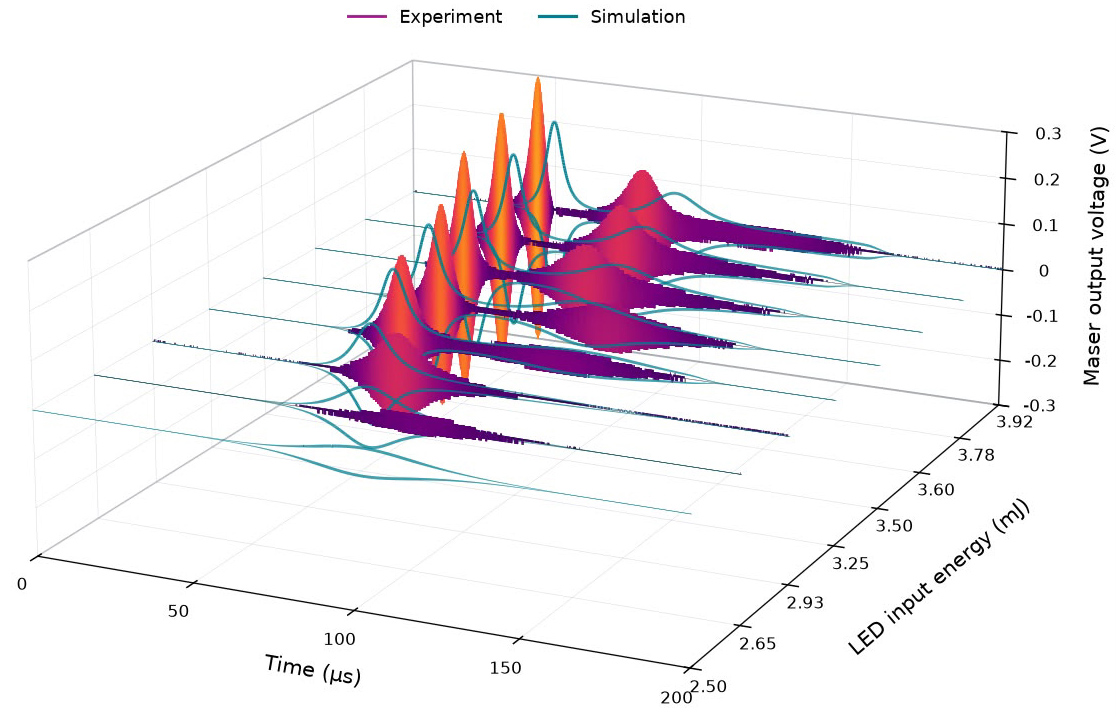}
\caption{Time profile of down-converted (to 70 MHz) maser bursts for different applied optical pump energies; the orange-pump traces are experimental data; the teal blue envelope is the prediction of our best-fitted quantum mechanical model. The vertical scale is the output voltage of our superheterodyne receiver, whose measured sensitivity (to a 1.45~GHz maser oscillation applied to its input) is
417~V/$\sqrt{\textrm{W}}$ or, equivalently, 13.3~V/$\sqrt{\textrm{mW}}$. The threshold for maser oscillation is here slightly lower than for the data shown in Fig.~\ref{fig:masing} due to a slightly different axial location (height) of the lollipop in the STO ring's bore.  
}
\label{fig:sim-exp_waterfal}
\end{figure}

\section{Quantum Mechanical Simulation}
The dynamics of the LED-pumped pentacene maser are modelled 
using a five-level Tavis–Cummings model\cite{Tavis1968ExactHamiltonian,Breeze2017Room-temperatureStates,Wu2020Room-TemperatureConcentrator} implemented in Julia using \verb=QuantumCumulants=\cite{kubo1962generalized,quantumcumulants}.
Optical excitation, spin-selective intersystem crossing, triplet decay, spin-lattice relaxation, pure dephasing,
and thermal cavity dissipation are incorporated through Lindblad operators evolving correlations (as elements of
a reduced density matrix) in a master-equation approach. 
Our model is the same as the simplest, mean-field version of what is described
in ref.~\onlinecite{carrera2026cavityqedsimulationmasermeanfield}, but with a few differences (described below);
its parameters are displayed in Table~\ref{tab:wide_table}.
All pentacene molecules are assumed to  initially occupy the singlet ground state,
while the cavity has an initial population of 4300 thermal photons. 
The resulting equations are integrated from 0 to 200 $\mu$s using the \verb=Rodas5P= solver.
Our model contains only two free parameters: $N$ and $F$, both discussed below.
The predicted instantaneous maser output power is fitted to measured experimental data
for 8 different optical excitation energies (each pump pulse lasts 150~$\mu$s) around the threshold for masing.
This data (in orange-purple) is shown along with the best-fitted model's predictions (in teal) in   Fig.~\ref{fig:sim-exp_waterfal}.

\begin{table*}[t!]
  \centering
  \begin{tabular}{|c|c|c|c|}
    \hline
Quantity & Symbol & Value & Notes \\
    \hline
X-Z transition frequency
  & \(f_{35} = \omega_{35} / ({2 \pi}) \) & \(\SI{1.4495}{\giga\hertz}\)
  & \\
 = TE$_{01\delta}$ mode frequency
  & $= f_\textrm{mode}$   &  &  experimentally tuned \\
loaded quality factor
  & \(Q_{\mathrm{loaded}}\) & \(6000\)
  & experimentally measured\\
mode dissipation rate
  & \(\kappa_\textrm{c} \equiv \omega_{35}/Q_{\mathrm{loaded}}\)
  & \(\SI{1.52e6}{\per\second}\) & \\
out-coupling
  & $S_{22}$
  & 0.3 & controls strength of output\\
mode volume ( of TE$_{01\delta}$ )  
  & $V_\textrm{mode}$ & 0.64~cm$^{3}$
  & ~2.1$\times$ larger than ref.~\onlinecite{Wu2020Room-TemperatureConcentrator}\\
spin--cavity coupling
  & \(g_\textrm{s} / (2 \pi) \) & \(\SI{0.16}{Hz}\)
  & ~1.4$\times$ smaller than ref.~\onlinecite{Wu2020Room-TemperatureConcentrator}\\
transverse relaxation time
  & \(T_2^*=2/\chi\) & \(\SI{1.82\,}{\micro\second}\)
  & \\
dephasing parameter  
  & \(\chi \equiv \kappa_\textrm{s} \) & \(\SI{1.1e6}{\per\second}\)  & \\
thermal cavity population
  & \(n_{\mathrm{th}}\) & \(4300\)
  & \\
\(S_1\rightarrow S_0\) & \(k_{\mathrm{sp}}\) & \(42.0\times10^6\)s$^{-1}$ & \\
\(S_1\rightarrow T_Z\) & \(k_{23}\) & \(5.52\times10^6\)s$^{-1}$ & \\
\(S_1\rightarrow T_Y\) & \(k_{24}\) & \(11.0\times10^6\)s$^{-1}$ & \\
\(S_1\rightarrow T_X\) & \(k_{25}\) & \(52.4\times10^6\)s$^{-1}$ & \\
\(T_Z\rightarrow S_0\) & \(k_{31}\) & \(0.2\times10^4\)s$^{-1}$ & \\
\(T_Y\rightarrow S_0\) & \(k_{41}\) & \(1.4\times10^4\)s$^{-1}$ & \\
\(T_X\rightarrow S_0\) & \(k_{51}\) & \(2.2\times10^4\)s$^{-1}$ & \\
\(T_Z\leftrightarrow T_X\) & \(k_{35}=k_{53}\) & \(1.1\times10^4\)s$^{-1}$ & \\
\(T_Z\leftrightarrow T_Y\) & \(k_{34}=k_{43}\) & \(2.8\times10^4\)s$^{-1}$ & \\
\(T_Y\leftrightarrow T_X\) & \(k_{45}=k_{54}\) & \(0.4\times10^4\)s$^{-1}$ & \\
optical-power conversion
  & $F$ & \(\SI{543}{\per\second\per\watt}\)
  & fitted parameter\\
number of participating spins
  & \(N\) & \(8.0\times10^{14}\)
  & fitted parameter\\
\bottomrule    
    \hline
  \end{tabular}
  \caption{Parameters used in quantum-mechanical simulation.}
  \label{tab:wide_table}
\end{table*}

Our model considers an ensemble of $N$ identical pentacene molecules,
whose X–Z transitions are coupled (all with the same common coupling coefficient
$g_\textrm{s}$) at zero detuning to a single microwave cavity mode at 1.4495~GHz. 
The time-dependent pumping (absorbed optical power versus time) provided by the LED is modelled using
an experimentally measured trace (volts versus time --recorded on a digital oscilloscope) from a
linear optical detector (Thorlabs DET~200), whose area under the curve is normalised to a total pulse energy of 3.3 mJ;
this latter value is itself inferred (via ray-tracing simulations --see Appendix E) from the value reported by a pyrolectric energy meter (Thorlabs ES245C).
This normalised trace is linearly interpolated in time and scaled in amplitude
to produce reasonably realistic pumping profiles for 8 different 
optical pulse energies ranging from 2.50 to 3.92~mJ.
The optical excitation rate per pentacene molecule, describing the transition from the singlet ground state to the excited singlet state, is calculated as
$F \times P_{\textrm{LED}}(t)$,
where $P_{\textrm{LED}}(t)$ is the instantaneous optical power and
$F$ is the effective optical-power-to-excitation-rate conversion coefficient
in units of s$^{-1}$W$^{-1}$ or, equivalently, J$^{-1}$.

The best-fit value for the number of pentacene molecules involved in masing is found
to be $N = 8 \times 10^{14}$ (\textit{cf.}~$N = 2.3 \times 10^{15}$ used in ref.~\onlinecite{carrera2026cavityqedsimulationmasermeanfield}).
It is informative to compare this number against the total number of pentacene molecules
dissolved inside the lollipop (taking into account the volume excluded by the spear head).
For a doping concentration of 0.1~\%, the latter is estimated\cite{ZikeChengMSc},
assuming no pentacene is lost during crystal growth,
to be $N_0 = 4 \times 10^{17}$ --a factor of 500 larger that our best-fit value for $N$. 
The extremely low fraction of pentacene molecules that actually participate in masing
has been noted by previous authors\cite{Wu2020Room-TemperatureConcentrator}. The extremely uneven
illumination of the crystal, as shown in Fig.~\ref{fig:invasive_vs_butt} can at least partially explain it.

The best-fit value for the optical-power-to-excitation-rate conversion coefficient is found to be
$F =543$~J$^{-1}$ (\textit{cf.}~$F = 3137$~J$^{-1}$ in ref.~\onlinecite{carrera2026cavityqedsimulationmasermeanfield}).
To very first approximation, in the case of an optically thin slice of ptc:ptp,
this coefficient should equal
$F_\textrm{t} = [\lambda_0^\textrm{pump}  / (hc)] \times \Lambda / A$, 
where $h c / \lambda_0^\textrm{pump}$ is the energy of a pump photon,
$\Lambda$ is pentacene's optical molecular cross-section,
and $A$ is the slice's area. 
Equating the latter to the area of the spear-head's
3 tear-drop facets and using our own measured value of 
$0.7 \times 10^{-17}$~cm$^2$ for pentacene's optical cross-section (at 532~nm)\cite{ZikeChengMSc}, 
we estimate $F_\textrm{t} \approx 25$~J$^{-1}$.
Given that, in the actual experiment (as Fig.~\ref{fig:invasive_vs_butt} again reveals),
the illumination of the lollipop's pentacene molecules is extremely non-uniform,
the factor of $\sim$22 discrepancy between this estimate and our best-fit value
for $F$ is not wholly unexpected.

In conclusion, we have demonstrated maser oscillation in a solid-state, room-temperature maser that is invasively pumped through a quartz waveguide by a single, chip-scale LED. We estimate, from experiment, the cooperativity of this maser to be $\approx 2$, and find that it can produce quasi-CW maser oscillations with features competitive with previously reported embodiments whose optical pump sources are substantially more voluminous and heavier. 
In our particular design, replacing the spear-headed lollipop by a butt-coupled
one causes the cooperativity to drop below unity, so precluding maser oscillation.
By combining FEM-based electromagnetic simulation, with optical ray tracing, and with cavity-QED modelling,
we can broadly explain our experimental results. In particular, we show how the maser's cooperativity can
be straightforwardly estimated as a Monte Carlo summation over ray-tracing data.  
Our analysis motivates further attempts to improve the illumination of the maser crystal towards
substantially improving performance.

\section{Acknowledgments}
V.U. thanks the Rank Prize\cite{RankPrizeUROP} for an Undergraduate Vacation Grant (2026) --enabling her to construct a more realistic optical ray-tracing model for this work. 
We thank Ben Gaskill of Gaskill Quartz Ltd., London, for manufacturing the STO ring resonator used in this work.
M.N. acknowledges the support of UK EPSRC grants: EP/R513052/1, EP/T51780X/1 and EP/V001914/1.
M.O. acknowledges the support of UK EPSRC grants: EP/V048430/1 and EP/Y00471X/1.

\section{Author Declarations}

\subsection{Conflict of Interest}
The authors declare no conflicts of interest.

\subsection{Author Contributions}

\textbf{Michael Newns}: writing - original draft; investigation - experimental (principal), data analysis (principal), writing - review \& editing (principal); investigation - electromagnetic simulation (principal); conceptualisation (supporting).
\textbf{Shirley Xu}: investigation - experimental (principal), data analysis (supporting); conceptualisation (supporting).
\textbf{Mingyang Liu}: investigation - optical (ray-tracing) simulation (principal); writing - review \& editing (supporting).
\textbf{Yifan Yu}: investigation - experimental (principal), data analysis (principal); writing - review \& editing (supporting).
\textbf{Zike Cheng}: graphical design - figures (lead); writing - review \& editing (supporting).
\textbf{Vanessa Ussalim}: investigation - optical (ray-tracing) simulation (principal); writing - review \& editing (supporting).
\textbf{Zitong Qin}: investigation - quantum mechanical (Julia) simulation (lead); writing - review \& editing (supporting).
\textbf{Ziqiu Huang}: investigation - experimental (supporting), writing - review \& editing (supporting).
\textbf{Max Attwood}: investigation - experimental (supporting); conceptualisation (supporting); resources (supporting).
\textbf{Mark Oxborrow}: resources/funding (lead); conceptualisation (lead); writing - review \& editing (principal); data analysis (supporting); investigation - experimental (supporting); investigation - electromagnetic, optical and quantum mechanical simulation (supporting).

\subsection{Data Availability}
The data that support the findings of this study are available
from the corresponding author upon reasonable request.

\section{Bibliography}

\bibliography{references2}

@article{Goldenberg1960AtomicMaser,
    title = {{Atomic hydrogen maser}},
    year = {1960},
    journal = {Physical Review Letters},
    author = {Goldenberg, Harold Mark and Kleppner, Daniel and Ramsey, Norman Foster},
    number = {8},
    pages = {361},
    volume = {5},
    publisher = {APS}
}

@article{Sherman2022Diamond-basedAmplifier,
  title={Diamond-based microwave quantum amplifier},
  author={Sherman, Alexander and Zgadzai, Oleg and Koren, Boaz and Peretz, Idan and Laster, Eyal and Blank, Aharon},
  journal={Science Advances},
  volume={8},
  number={49},
  pages={eade6527},
  year={2022},
  publisher={American Association for the Advancement of Science}
}

@article{Breeze2015EnhancedMasers,
    title = {{Enhanced magnetic Purcell effect in room-temperature masers}},
    year = {2015},
    journal = {Nature Communications},
    author = {Breeze, Jonathan and Tan, Ke Jie and Richards, Benjamin and Sathian, Juna and Oxborrow, Mark and Alford, Neil Mc N.},
    number = {1},
    month = {2},
    pages = {1--6},
    volume = {6},
    publisher = {Nature Publishing Group},
    url = {https://www.nature.com/articles/ncomms7215},
    doi = {10.1038/ncomms7215},
    issn = {2041-1723}
}

@article{Tavis1968ExactHamiltonian,
    title = {{Exact solution for an N-molecule radiation-field Hamiltonian}},
    year = {1968},
    journal = {Phys. Rev.},
    author = {Tavis, Michael and Cummings, Frederick W.},
    number = {2},
    pages = {379--384},
    volume = {170},
    doi = {10.1103/physrev.170.379},
    issn = {0031899X}
}

@article{Wu2020InvasiveCoupling,
    title = {{Invasive optical pumping for room-temperature masers, time-resolved EPR, triplet-DNP, and quantum engines exploiting strong coupling}},
    year = {2020},
    journal = {Optics Express},
    author = {Wu, Hao and Wu, Hao and Mirkhanov, Shamil and Ng, Wern and Chen, Kuan-Cheng and Xiong, Yuling and Xiong, Yuling and Oxborrow, Mark},
    number = {20},
    month = {9},
    pages = {29691--29702},
    volume = {28},
    publisher = {Optica Publishing Group},
    url = {https://opg.optica.org/viewmedia.cfm?uri=oe-28-20-29691&seq=0&html=true https://opg.optica.org/abstract.cfm?uri=oe-28-20-29691 https://opg.optica.org/oe/abstract.cfm?uri=oe-28-20-29691},
    doi = {10.1364/OE.401294},
    issn = {1094-4087},
    pmid = {33114862}
}

@article{Sathian2025LED-pumpedMaser,
    title = {{LED-pumped room-temperature solid-state maser}},
    year = {2025},
    journal = {Communications Engineering },
volume = {4},
page = {122},
    author = {Long, Sophia and Lopez, Lisa and Ford, Bethan and Balembois, François and Montis, Riccardo and Ng, Wern and Arroo, Daan and Alford, Neil and Torun, Hamdi and Sathian, Juna},
    month = {7}
}

@misc{Oxborrow2012MaserAssembly,
    title = {{Maser assembly}},
    year = {2012},
    author = {Oxborrow, Mark and {US patent US9608396B2}},
    month = {8},
    url = {http://hdl.handle.net/2433/64950.},
    institution = {NPL}
}

@article{Ng2023MoveElectrodynamics,
    title = {{Move Aside Pentacene: Diazapentacene-Doped para-Terphenyl, a Zero-Field Room-Temperature Maser with Strong Coupling for Cavity Quantum Electrodynamics}},
    year = {2023},
    journal = {Advanced Materials},
    author = {Ng, Wern and Xu, Xiaotian and Attwood, Max and Wu, Hao and Meng, Zhu and Chen, Xi and Oxborrow, Mark},
    number = {22},
    volume = {35},
    month = {6},
    pages = {2300441},
    publisher = {John Wiley and Sons Inc}
}

@article{Salvadori2017NanosecondMASER,
    title = {{Nanosecond time-resolved characterization of a pentacene-based room-temperature MASER}},
    year = {2017},
    journal = {Scientific Reports},
    author = {Salvadori, Enrico and Breeze, Jonathan D. and Tan, Ke Jie and Sathian, Juna and Richards, Benjamin and Fung, Mei Wai and Wolfowicz, Gary and Oxborrow, Mark and Alford, Neil Mc N. and Kay, Christopher W.M.},
    number = {1},
    month = {2},
    pages = {1--8},
    volume = {7},
    publisher = {Nature Publishing Group},
    url = {https://www.nature.com/articles/srep41836},
    doi = {10.1038/srep41836},
    issn = {2045-2322},
    pmid = {28169331}
}

@article{Sherman2021PerformanceMasers,
    title = {{Performance analysis of diamond-based masers}},
    year = {2021},
    journal = {Journal of Applied Physics},
    author = {Sherman, Alexander and Buchbinder, Lotem and Ding, Siyuan and Blank, Aharon},
    number = {14},
    month = {4},
    pages = {144503},
    volume = {129},
    publisher = {American Institute of Physics Inc.},
    url = {/aip/jap/article/129/14/144503/1078785/Performance-analysis-of-diamond-based-masers},
    doi = {10.1063/5.0045802/1078785},
    issn = {10897550}
}

@article{Arroo2021PerspectiveMasers,
    title = {{Perspective on room-temperature solid-state masers}},
    year = {2021},
    journal = {Applied Physics Letters},
    author = {Arroo, Daan M. and Alford, Neil Mc N. and Breeze, Jonathan D.},
    number = {14},
    month = {10},
    pages = {140502},
    volume = {119},
    publisher = {American Institute of Physics Inc.},
    url = {/aip/apl/article/119/14/140502/40745/Perspective-on-room-temperature-solid-state-masers},
    doi = {10.1063/5.0061330/40745},
    issn = {00036951}
}

@article{Breeze2017Room-temperatureStates,
    title = {{Room-temperature cavity quantum electrodynamics with strongly coupled Dicke states}},
    year = {2017},
    journal = {npj Quantum Information},
    author = {Breeze, Jonathan D. and Salvadori, Enrico and Sathian, Juna and McN. Alford, Neil and Kay, Christopher W.M.},
    number = {1},
    volume = {3},
    publisher = {Nature Research}
}

@article{Wu2020Room-TemperatureConcentrator,
    title = {{Room-Temperature Quasi-Continuous-Wave Pentacene Maser Pumped by an Invasive \texttt{Ce:YAG} Luminescent Concentrator}},
    year = {2020},
    journal = {Physical Review Applied},
    author = {Wu, Hao and Xie, Xiangyu and Ng, Wern and Mehanna, Seif and Li, Yingxu and Attwood, Max and Oxborrow, Mark},
    number = {6},
    month = {12},
    volume = {14},
    publisher = {American Physical Society}
}

@article{Oxborrow2012Room-temperatureMaser,
    title = {{Room-temperature solid-state maser}},
    year = {2012},
    journal = {Nature},
    author = {Oxborrow, Mark and Breeze, Jonathan D. and Alford, Neil M.},
    number = {7411},
    month = {8},
    pages = {353--356},
    volume = {488},
    publisher = {Nature Publishing Group},
    url = {https://www.nature.com/articles/nature11339},
    doi = {10.1038/nature11339},
    issn = {1476-4687}
}

@article{Wu2019UnravelingField,
    title = {{Unraveling the Room-Temperature Spin Dynamics of Photoexcited Pentacene in Its Lowest Triplet State at Zero Field}},
    year = {2019},
    journal = {Journal of Physical Chemistry C},
    author = {Wu, Hao and Ng, Wern and Mirkhanov, Shamil and Amirzhan, Arman and Nitnara, Supamas and Oxborrow, Mark},
    number = {39},
    month = {10},
    pages = {24275--24279},
    volume = {123},
    publisher = {American Chemical Society},
    url = {https://pubs.acs.org/sharingguidelines},
    doi = {10.1021/ACS.JPCC.9B08439/SUPPL{\_}FILE/JP9B08439{\_}SI{\_}001.PDF},
    issn = {19327455}
}

@article{Takeda2002Zero-fieldCrossing,
    title = {{Zero-field electron spin resonance and theoretical studies of light penetration into single crystal and polycrystalline material doped with molecules photoexcitable to the triplet state via intersystem crossing}},
    year = {2002},
    journal = {Journal of Chemical Physics},
    author = {Takeda, Kazuyuki and Takegoshi, K. and Terao, Takehiko},
    number = {10},
    month = {9},
    pages = {4940--4946},
    volume = {117},
    publisher = {AIP Publishing},
    url = {/aip/jcp/article/117/10/4940/459543/Zero-field-electron-spin-resonance-and-theoretical},
    doi = {10.1063/1.1499124},
    issn = {0021-9606}
}

@article{Ng2024Maser-in-a-shoebox:Field,
    title = {{`Maser-in-a-shoebox': A portable plug-and-play maser device at room temperature and zero magnetic field}},
    year = {2024},
    journal = {Applied Physics Letters},
    author = {Ng, Wern and Wen, Yongqiang and Attwood, Max and Jones, Daniel C. and Oxborrow, Mark and Alford, Neil Mc N. and Arroo, Daan M.},
    number = {4},
    month = {1},
    pages = {44004},
    volume = {124},
    publisher = {American Institute of Physics Inc.},
    url = {/aip/apl/article/124/4/044004/3061570/Maser-in-a-shoebox-A-portable-plug-and-play-maser},
    doi = {10.1063/5.0181318/3061570},
    issn = {00036951},
    arxivId = {2310.09269}
}

@article{wu_quantum_sensing2022,
author = {Hao Wu  and Shuo Yang  and Mark Oxborrow  and Min Jiang  and Qing Zhao  and Dmitry Budker  and Bo Zhang  and Jiangfeng Du },
title = {Enhanced quantum sensing with room-temperature solid-state masers},
journal = {Science Advances},
volume = {8},
number = {48},
pages = {eade1613},
year = {2022},
doi = {10.1126/sciadv.ade1613}
}

@article{varshavsky2025solid,
  title={Solid-State Maser with Microwatt Output Power at Moderate Cryogenic Temperatures},
  author={Varshavsky, Yefim and Zgadzai, Oleg and Blank, Aharon},
  journal={arXiv preprint arXiv:2504.06846},
  year={2025}
}

@book{clauss2008ruby,
  title={Ruby masers},
  author={Clauss, Robert C and Shell, James S},
  journal={Low-Noise Systems in the Deep Space Network},
editor={Reid, Macgregor S.},
publisher={JPL Deep Space Communications And Navigation Series, Jet Propulsion Laboratory},
  year={2008}
}

@phdthesis{Farrell2008,
  author       = {Daniel J. Farrell},
  title        = {Characterising the Performance of Luminescent Solar Concentrators},
  school       = {Imperial College London},
  year         = {2008}
}

@misc{FarrellPVTraceOriginal,
  author       = {Daniel J. Farrell},
  title        = {{pvtrace: Optical ray tracing for luminescent materials and spectral converter photovoltaic devices}},
  howpublished = {\url{https://github.com/danieljfarrell/pvtrace}},
  urldate      = {2025-11-03}
}

@article{verma2023ray,
  title={Ray-trace modeling to characterize efficiency of unconventional luminescent solar concentrator geometries},
  author={Verma, Shomik and Farrell, Daniel J and Evans, Rachel C},
  journal={ACS Applied Optical Materials},
  volume={1},
  number={5},
  pages={1012--1025},
  year={2023},
  publisher={ACS Publications}
}

@misc{PVTraceModified,
  author       = {{Shomik Verma}},
  title        = {{Pvtrace-Sv}},
  howpublished = {\url{https://github.com/shomikverma/pvtrace-sv}},
  urldate      = {2025-11-03}
}

@article{nelson1981laser,
  title={Laser-induced phonon spectroscopy. Optical generation of ultrasonic waves and investigation of electronic excited-state interactions in solids},
  author={Nelson, Keith A and Lutz, DR and Fayer, MD and Madison, Larry},
  journal={Physical Review B},
  volume={24},
  number={6},
  pages={3261},
  year={1981},
  publisher={APS}
}

@manual{osmar,
  key        = {\texttt{OSRAM OSTAR™ Projection Power, LE CG P2AQ}},
  year         = {2025},
  url            = {ams-OSRAM AG},
  organization          = {https://ams-osram.com/products/leds/color-leds/osram-osram-ostar-projection-power-le-cg-p2aq},
}

@book{twyman88,
  title={Prism and Lens Making, Second Edition: A Textbook for Optical Glassworkers},
  author={Twyman, F},
  year={1988},
  publisher={Taylor and Francis}
}

@phdthesis{supamas_2020,
    author = {Nitnara, Supamas},
    title = {A Comparison of The Glass Former 1,3,5-Tri(Naphthyl)Benzene and The Crystalline p-Terphenyl Hosts for Pentacene Masers},
    school = {Imperial College London, Department of Materials},
    year = {2020}
}

@article{STO_diel,
  author={Zhao, Zijun C. and Goryachev, Maxim and Krupka, Jerzy and Tobar, Michael E.},
  journal={IEEE Transactions on Ultrasonics, Ferroelectrics, and Frequency Control}, 
  title={Precision Multi-Mode Dielectric Characterization of a Crystalline Perovskite Enables Determination of the Temperature-Dependent Phase Transitions}, 
  year={2022},
  volume={69},
  number={1},
  pages={423-429},
  doi={10.1109/TUFFC.2021.3108118}}

@misc{carrera2026cavityqedsimulationmasermeanfield,
      title={Cavity-QED Simulation of a Maser beyond the Mean-Field Approximation}, 
      author={Niall Randall Carrera and Yining Jiang and Xinpeng Shu and Hao Wu and Mark Oxborrow},
      year={2026},
      eprint={2412.21166},
      archivePrefix={arXiv},
      primaryClass={quant-ph},
      url={https://arxiv.org/abs/2412.21166}, 
}

@book{Poole1983,
    booktitle = {Electron spin resonance: a comprehensive treatise on experimental techniques, Second Ed.},
    author = {Poole, Charles P. },
    publisher = {Wiley},
    year = {1983}
}

@article{bogle1959zero,
  title={Zero-field masers},
  author={Bogle, GS and Symmons, HF},
  journal={Australian Journal of Physics, vol. 12, p. 1},
  volume={12},
  pages={1},
  year={1959}
}

@book{A.E.Siegman1964MicrowaveMasers,
    title = {{Microwave Solid-state Masers}},
    year = {1964},
    author = {{A. E. Siegman}},
    publisher = {McGraw-Hill electrical and electronic engineering series},
    url = {https://books.google.co.uk/books/about/Microwave_Solid_state_Masers.html?id=t9ZEAAAAIAAJ&redir_esc=y}
}

@article{xiang2025all,
  title={All-metallic magnetic Purcell enhancement in a thermally stable room-temperature maser},
  author={Xiang, Rongrong and Bugnon, Philippe and Moghaddam, Maliheh Khatibi and Fleury, Romain},
  journal={Nature Communications},
  volume={16},
  pages={11214},
  year={2025}
}

@misc{RankPrizeUROP,
  howpublished = {Rank Prize, \url{https://www.rankprize.org/}}
}

@misc{ZikeChengMSc,
howpublished = {Zike Cheng, MSc thesis, Dept.~Materials, Imperial College London (2025)}
}

@article{kubo1962generalized,
  title={Generalized cumulant expansion method},
  author={Kubo, Ryogo},
  journal={Journal of the Physical Society of Japan},
  volume={17},
  number={7},
  pages={1100--1120},
  year={1962},
  publisher={The Physical Society of Japan}
}

@article{quantumcumulants,
  doi = {10.22331/q-2022-01-04-617},
  url = {https://doi.org/10.22331/q-2022-01-04-617},
  title = {Quantum{C}umulants.jl: {A} {J}ulia framework for generalized mean-field equations in open quantum systems},
  author = {Plankensteiner, David and Hotter, Christoph and Ritsch, Helmut},
  journal = {{Quantum}},
  issn = {2521-327X},
  publisher = {{Verein zur F{\"{o}}rderung des Open Access Publizierens in den Quantenwissenschaften}},
  volume = {6},
  pages = {617},
  month = jan,
  year = {2022}
}

\renewcommand\thesection{APPENDIX~\Roman{section}} 
\setcounter{section}{0}

\renewcommand\thefigure{A\arabic{figure}} 
\setcounter{figure}{0} 

\renewcommand\thetable{S\arabic{table}} 
\setcounter{table}{0} 

\maketitle

\appendix

\section{RF simulation}

RF modelling is conducted in COMSOL Multiphysics (version 6.0), using the RF module to conduct eigenfrequency analysis of a simplified model of the cavity internals. The STO resonator is modelled as a lossless dielectric with $\epsilon_\textrm{r}=316.3$ (from ref.~\onlinecite{STO_diel}), supported by a PTFE stand (internal material parameters), with perfect-electric-conductor boundary conditions simulating the copper cavity's walls. No attempt is made to electromagnetically simulate the ptc:ptp crystal, waveguide or coupling loops, as these are non-magnetic and not found to significantly affect the mode's shape. The simulated data set is used to extract a high resolution interpolation of
$|\textbf{B}|^2 = \mu_0^2 |\textbf{H}|^2$ in the central bore of the dielectric resonator.

\section{Optical Ray-tracing Simulation}

Within our simulation, we treat the LED as a flat rectangular lambertian light source, emitting through a layer of optical coupling liquid into the waveguide. Light from the LED that is successfully captured by the waveguide then propagates along it by total internal reflection until it meets the waveguide's terminating ptc:ptp crystal.  
In our simulation, we assume that 0.1\%~ptc:ptp is a linear (bleaching effects are ignored) optical  absorber obeying the Beer-Lambert law in the direction of optical transport.
The simulation utilises the main ray-tracing functions defined in \verb=photon_tracer.py= of the original PVTrace repository but with a few modifications and customisations/simplifications. Our system comprises 3 ``nodes'' surrounded by air (refrective index = 1):  
%
%
%

[1] The ptc:ptp crystal (refractive index = 1.63) takes the geometry of a \verb=Cylinder()=, with a base radius of 6~mm, a height of 13.34~mm, and with the node’s position at the origin and shifted -0.01~mm in the z direction,  The ptc:ptp crystal’s Beer-Lambert linear absorption coefficient \verb=alpha= (in units of neper mm$^{-1}$) is represented as a spectrum function, similar to the \verb=absorption()= function defined for luminescent concentrators in the original PVTrace repository. The spectrum is made by linearly interpolating the data values derived from our own UV-Vis measurements (the alpha coefficient and their respective wavelengths) using \verb=np.interp()=. During the simulation run, the spectrum is passed as a \verb=Distribution= object. 

[2] The waveguide (refractive index = 1.46) takes the geometry defined by an STL file, scaled to mm. This node’s position (origin in the STL file → origin in the ray-tracing scene) is then similarly shifted -0.01~mm in the z direction. The waveguide is non-absorbing and non-emitting.

[3] The light source is defined as a \verb=rectangular mask()= with dimensions 3.2~$\times$~2.6~mm, having a light direction of a Lambertian cone with a half-angle of 36.4$^\circ$. The position of the light source is the waveguide’s base STL coordinates relative to the origin (0, 0, -117)~mm, without any shift in the z direction. This means that the light source is embedded inside the waveguide, +0.01~mm in the z-direction from the waveguide’s bottom face. The emission wavelengths are represented by a spectrum function, similar to the \verb=emission()= function defined for a luminescent concentrator in the original PVTrace repository. The spectrum is made by interpolating data values (for relative emission intensity versus wavelength) using \verb=CubicSpline()=. During the simulation run, this spectrum is passed as a \verb=Distribution= object. 

The simulation is then run with 10,000 rays emitting from the light source. The ray-tracing run tracks each ray from one interface to the next. As each ray travels between interfaces, Beer-Lambert sampling determines whether it gets absorbed along its path through the ptc:ptp crystal. If the rays are absorbed by the ptc:ptp crystal, a non-radiative absorption event occurs, terminating the ray. Otherwise, the rays continue to the next interface, where Fresnel reflection or transmission is resolved. The ray can also be terminated if it leaves the \verb=World Sphere= or encounters more than 1000 interface interactions.

\section{Waveguide Shaping}

Our two waveguides were hand-made using traditional prism-making methods along the lines of ref.~\onlinecite{twyman88}.
A section of 5-mm-in-diameter fused quartz rod is scored (with a diamond scribe) and snapped to length (slightly over 13 cm).
Polished end faces and/or facets are then formed on this length by mounting it with blocking wax (= blend of 85\% colophony, 15\% bees wax by weight) into jigs incorporating glass ``waster'' surfaces.
A facet in the making and its surrounding (angle-defining) waster surface were pressed by hand against the surface of a flat rotating grinding/polishing wheel, onto which progressively finer grades of abrasive paper were attached. The final polishing of each face/facet was done against a napped cloth impregnated with 3-$\mu$m diamond paste on a similar wheel. On their input ends, both waveguides received a polished flat face (cut square to the rod's axis) to contact and cover the pump LED's flat outputting face. The `butt-coupled' waveguide received an identical flat face at its other, outputting end. The outputting end of the spear-headed waveguide in contrast comprised three flat faces (facets), like the apex of a triangular pyramid (tetrahedron), but with a elliptical edge where each of these facets met the waveguide's cylindrical shank (see main text, Fig.~\ref{fig:anatomy}). The waveguide's length helps isolate the maser crystal from transient magnetic fields generated by pulsed electrical current flowing through the LED; the length could certainly have been reduced by placing one or several layers of magnetic (mumetal) shielding between the LED and maser crystal, but this sophistication was not implemented.

\section{LED Driving}

 The LED is driven by a home-made circuit that discharges a capacitor (\textit{viz.}~an EVOX RIFA 22,000 $\mu$F, 63~V, low-ESR electrolytic) through a low-on-resistance MOSFET switch (comprising four IRFB3077PbF transistors wired in parallel, whose gates are driven by the same single high-current TC4452VAT driver, which itself is controlled by a standard TTL) into the LED. The voltage across a low-inductance 0.02-$\Omega$ resistor, placed in series between four MOSFETs' source electrodes and ground, is used to monitor the current flowing through the LED.
 As a precaution against overheating during sustained pulsed operation, the LED module is mounted with thermally-conductive paste atop an aluminium heatsink. The highest drive currents used in this work exceed the maximum surge current (14 A) stated in the LED's datasheet~\cite{osmar} by a factor of 2. The risk of damage is mitigated by restricting the current pulse's duration to 150 $\mu$s and keeping to a low duty cycle of 1 pulse per second. These measures allowed for hours of pulsed above-threshold maser without LED failure (or even any noticeable degradation). 

\section{Optical Pump Calibration}
 
The time-integrated total optical energy outputted by the LED over a single pump pulse can be controlled by varying either (i) the duration of the TTL pulse controlling how long the LED-driver's MOSFET switch stays open or (ii) the voltage to which the driver's capacitor is charged. For all of the experiments reported in this paper, we fix the former at 150~$\mu$s and vary the latter. 
The pulse energy actually received by the maser crystal is determined through a combination of
experimental measurements and our most advanced polychromatic optical ray-tracing simulation.
A ``dummy'' waveguide is first made from a 13-cm length of the same 5-mm fused quartz rod with end faces cut at right angles (= ``cut square'') to the rod's axis and polished.
The rod is oriented vertically with it bottom end face sat onto the (upwardly facing) LED, with
a drop of Santovac 5 inserted between them (this drop gets immediately squeezed, by the waveguide's weight, into a thin film). 
A silicon photo-diode (\textit{viz.} Thorlabs DET200; modern equivalent: DET100A2) is positioned facing and a few cm away from the LED (the exact orientation and location is seldom important since the LED is bright and the detector has plenty of dynamic range). This diode is used as a linear, uncalibrated detector to measure the time profile of the LED's optical output.

\begin{figure}
    \centering
    \includegraphics[width=0.9\linewidth]{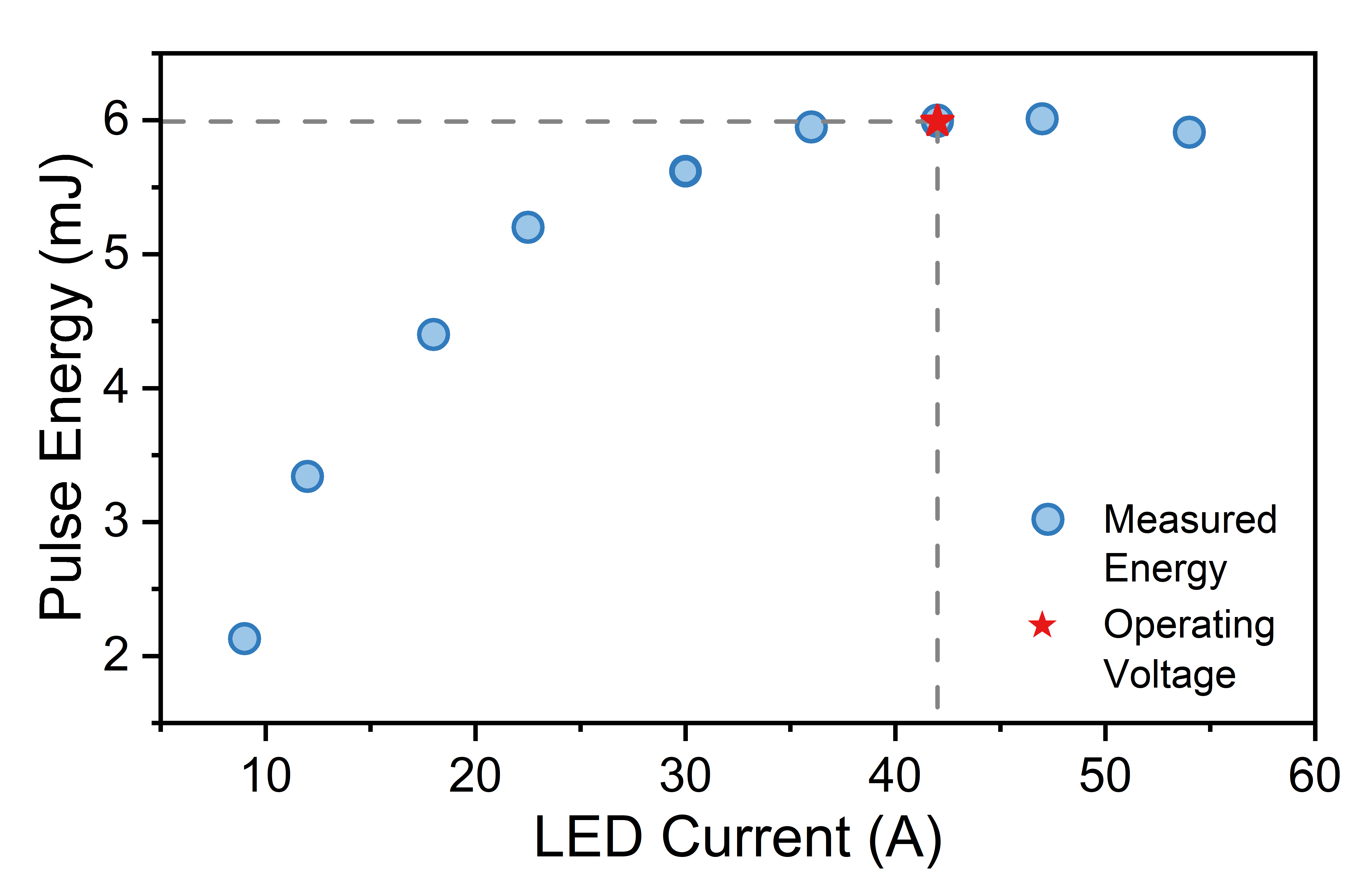}
    \caption{Measured LED output energy vs peak LED drive current for 150-$\mu$s pulses. }
    \label{fig:LED_cal}
\end{figure}

Light originating from the LED that escapes upwards, into air, from the dummy waveguide's upper outputting end face
is received by a downward-facing Thorlabs ES245C pyroelectric energy sensor placed just above the end face.
As shown in Fig.~\ref{fig:LED_cal}, the energy recorded by this sensor depends on the voltage to which the drive capacitor is charged,
which in turn determines the peak current that flows through the LED.
One sees that, above a peak drive current of 30~A or so  (14~A is the spec.~sheet's maximum allowable current),
one reaches diminishing returns. 
Each time profile recorded by the DET-200 is then converted/rescaled to units of power (watts) 
by demanding that the area-under-the-curve equal the measured total pulse energy reported by the energy meter. This allows for the instantaneous outputted optical power as a function of time to be determined, as shown at the bottom of Fig.~\ref{fig:masing} (main text) for two pulse energies (thus two different voltages on the LED-driver's capacitor). 

The measured optical energy/power outputted into air from a square-cut end face of a 5-mm quartz rod does not equate directly
to what gets absorbed by a ptc:ptp maser crystal (whose refractive index differs considerably from air's) grown onto the end of a similar 5-mm quartz rod (whose end is either spear-shaped on a flat square-cut). 

Using our most advance polychromatic ray tracing we find that, for 10,000 rays emitted by the LED at the bottom of 5-mm quartz waveguide:
(a) 9560 rays exit the waveguide's square-cut upper end-face into air and thereupon impinge on the surface of the ES245C energy meter;
(b) 9276 rays get absorbed by a ptc:ptp crystal atop a spear-headed waveguide as per the left side of 
Fig.~\ref{fig:cross_section};
(c) 9317 rays get absorbed by a ptc:ptp crystal butt-coupled against the waveguide's square-cut upper end-face as per the right side of Fig.~\ref{fig:cross_section};
Thus, the optical energy received by the ptc:ptp crystal atop the spear-headed or flat-top waveguides equals $9276/9560 \approx 0.970$
or $9317/9560 \approx 0.975$ times the energy reported by the ES245C energy sensor, respectively.

\end{document}